\documentclass[aps,prb,10pt,twocolumn,notitlepage,superscriptaddress]{revtex4-2}
\usepackage{amsmath,amssymb,graphicx,leftindex}
\usepackage[caption=false]{subfig}

\usepackage{xcolor}
\usepackage{bm}
\usepackage{comment}
\newcommand{\wIn}{\omega_{\textrm{in}}}
\newcommand{\wIO}{\omega_{\textrm{i0}}}
\newcommand{\wIC}{\omega_{\textrm{ic}}}
\newcommand{\TimeOrder}{\mathcal{T}}

\newcommand{\expectNormalOrder}[1]{\langle : \hspace{-0.065cm} #1 \hspace{-0.065cm} : \rangle}

\newcommand{\Eminus}{\mathcal{E}^{-}}
\newcommand{\Eplus}{\mathcal{E}^{+}}

\newcommand{\aVec}{\OP{\boldsymbol{a}}{}}
\newcommand{\aVecD}{\OPc{\boldsymbol{a}}{}}
\newcommand{\bBold}{\bm{b}}
\newcommand{\bVecX}[1]{\OP{\bBold}{#1}}
\newcommand{\bVecXD}[1]{\OPc{\bBold}{#1}}
\newcommand{\bInVec}{\OP{\bBold}{\textrm{in}}}

\newcommand{\bOutVec}{\OP{\bBold}{\textrm{out}}}
\newcommand{\bOutVecD}{\OPc{\bBold}{\textrm{out}}}
\newcommand{\bInVecX}[1]{\OP{\bBold}{\textrm{in},#1}}
\newcommand{\bOutVecX}[1]{\OP{\bBold}{\textrm{out},#1}}
\newcommand{\bInVecXD}[1]{\OPc{\bBold}{\textrm{in},#1}}
\newcommand{\bOutVecXD}[1]{\OPc{\bBold}{\textrm{out},#1}}
\newcommand{\Dop}{\hat{\bm{D}}}
\newcommand{\Uop}{\hat{\bm{U}}}
\newcommand{\fVec}[1]{\vec{\boldsymbol{f}}^{(#1)}}
\newcommand{\FVecIJ}[2]{\vec{\boldsymbol{F}}^{(#2)}_{#1}}

\newcommand{\polVec}{\vec{\mathbf{e}}}
\newcommand{\polVecC}{\vec{\mathbf{e}}^\dag}
\newcommand{\polTensor}{\vec{\boldsymbol{\epsilon}}}
\newcommand{\polTensorC}{\vec{\boldsymbol{\epsilon}}^\dag}

\newcommand{\polKetBra}[2]{\ketbra{\mathbf{e}^{#1}}{\mathbf{e}^{#2}}}

\newcommand{\seeAppendixX}[1]{[see Appendix #1]}

\newcommand{\Ham}{\hat{H}}
\newcommand{\ketbra}[2]{\left|#1\middle>\middle<#2\right|}
\newcommand{\braket}[2]{\left<#1\middle|#2\right>}

\newcommand{\bra}[1]{\left<#1\right|}
\newcommand{\ket}[1]{\left|#1\right>}
\newcommand{\OPc}[2]{\hat{#1}_{#2}^{\dag}}
\newcommand{\OP}[2]{\hat{#1}_{#2}^{\vphantom{\dag}}}
\newcommand{\CD}[1]{\OPc{c}{#1}}
\newcommand{\C}[1]{\OP{c}{#1}}
\newcommand{\A}[1]{\OP{a}{#1}}
\newcommand{\AD}[1]{\OPc{a}{#1}}
\newcommand{\BD}[1]{\OPc{b}{#1}}
\newcommand{\B}[1]{\OP{b}{#1}}
\newcommand{\ND}[1]{\hat{n}_{#1}}

\DeclareMathOperator{\Real}{Re}
\DeclareMathOperator{\Imag}{Im}

\newcommand{\hc}{\textrm{h.c.}}

\newcommand{\captiontitle}[1]{\textbf{#1}}

\newcommand{\expect}[1]{\langle #1 \rangle}

\newcommand{\BinXD}[1]{\OPc{b}{\textrm{in},#1}}
\newcommand{\BinX}[1]{\OP{b}{\textrm{in},#1}}
\newcommand{\BoutXD}[1]{\OPc{b}{\textrm{out},#1}}
\newcommand{\BoutX}[1]{\OP{b}{\textrm{out},#1}}

\begin{document}
\title{Many-Body Photon Blockade and Quantum Light Generation from \\Cavity Quantum Materials}

\author{Benjamin Kass}
\affiliation{Department of Physics and Astronomy, University of Pennsylvania, Philadelphia, PA 19104, USA}
\author{Spenser Talkington}
\affiliation{Department of Physics and Astronomy, University of Pennsylvania, Philadelphia, PA 19104, USA}
\author{Ajit Srivastava}
\affiliation{Department of Physics, Emory University, Atlanta, GA 30322, USA}
\affiliation{Department of Quantum Matter Physics, University of Geneva, Geneva 1211, Switzerland}
\author{Martin Claassen}
\affiliation{Department of Physics and Astronomy, University of Pennsylvania, Philadelphia, PA 19104, USA}
\date{November 13, 2024}
\begin{abstract}
The strong coupling regime of photons and quantum materials inside optical cavities has emerged as a promising environment for manipulating states of matter with light. Here, in turn, we show that photons bear witness to cavity quantum-electrodynamical modifications of the material, leading to profoundly non-classical properties of light passing through the cavity. By generalizing quantum-optical input-output relations to correlated quantum materials, we study the second-order photon coherence $g^{(2)}(t)$ and demonstrate that antibunching of transmitted photons serves as direct evidence of light-induced changes to the cavity-embedded material. We show that materials near a quantum critical point can realize a collective many-body photon blockade, enabling the generation of single photons or Einstein-Podolsky-Rosen pairs via leveraging strong matter fluctuations. Our findings provide new routes for interrogating and harnessing cavity-embedded quantum materials as quantum light sources, as a resource for photon-based computation and quantum sensing.
\end{abstract}

\maketitle

\begin{figure*}[hbt!]
    \centering
    \includegraphics[width=0.9\textwidth]{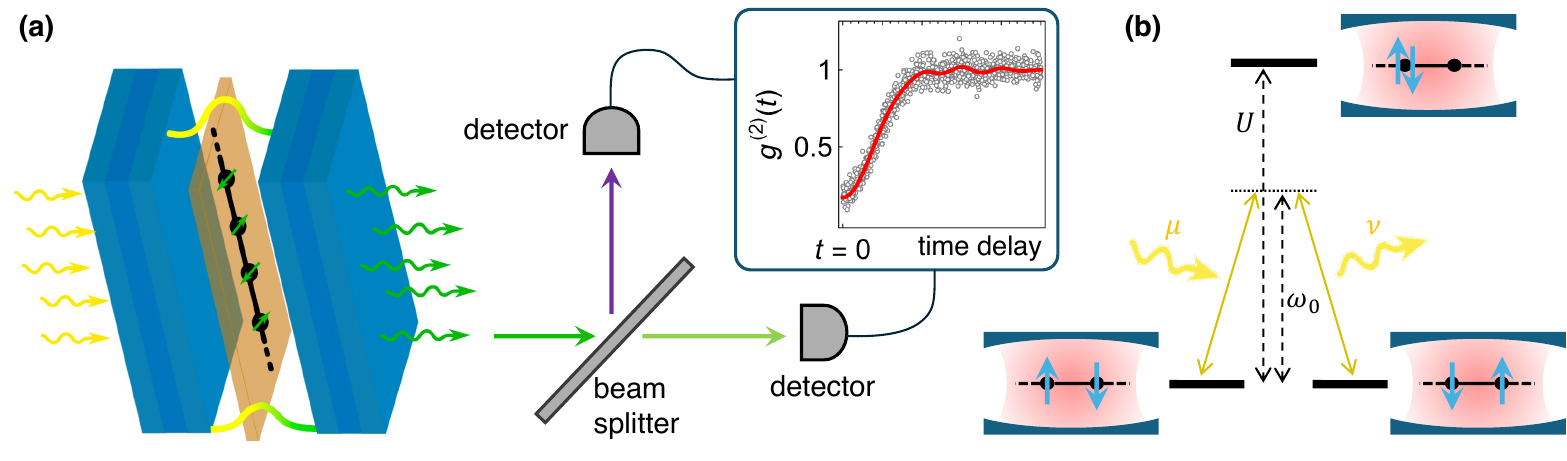}
    \caption{\textbf{Photon Coincidence Spectroscopy of Cavity Quantum Materials.} (a) Schematics of the strong light-matter coupling regime of quantum materials close to a QCP, placed inside a high finesse cavity. Seeding the cavity with photons can push the material across a phase transition, which is reflected in pronounced (anti)bunching of photons transmitted through the cavity, measurable via the higher-order photon coherence $g^{(2)}(t)$. (b) Strong coupling between light and elementary magnetic excitations in Mott insulators is mediated via repeated Raman processes while the photon resides in the cavity.}
    \label{fig:Schematics}
\end{figure*}

\section{Introduction}

The strong light-matter coupling regime of photons and correlated quantum materials has recently garnered much attention \cite{bloch22,schlawin21,hubener21,hubener24,garciavidal21}. In conventional quantum-optical settings, individual localized emitters become strongly coupled to photon modes inside optical cavities, generating polariton excitations with hybrid light and matter characteristics that are a rich resource for quantum information processing \cite{raimond2001manipulating,deng10,chang2014quantum,blais20,blais21,noh17}. Recent studies have started to bridge the fields of cavity quantum electrodynamics (cQED) and correlated electrons, by demonstrating strong coupling of photons to the collective modes of many-body systems in cavities \cite{basov20,soykal10,smolka14,maissen14,zhang16,halbhuber20,keller20,mueller20}, and exploring whether photon fluctuations can suffice to alter ground state properties of quantum materials \cite{jarc23,orgiu15,ebbesen16,nagarajan21, schlawin18,sentef18,curtis19,grankin21,juraschek19,ashida20,ashida20b,latini21,mazza19,kiffner19,sentef20,curtis22, flores2023non, chiocchetta21,bostrom22,masuki23,sentef19,shaffer23,eckhardt22,weber23,helmrich2024ultrastrong,kipp2024}. However, monitoring the \textit{quantum} electrodynamical dressing of states of matter in cavities, and distinguishing these photon-induced changes from more mundane screening or interface effects due to the cavity remains a central challenge. At the same time, observable ramifications for the state of \textit{light} remain largely unexplored, raising intriguing questions regarding whether correlated electron systems can be harnessed as sources of quantum photon states, with applications to quantum computing and metrology.

The starting point for this work rests on the observation that, if strong light-matter coupling indeed alters the collective phase of a quantum material, then photons must necessarily be witnesses to these changes through their entanglement with the material \cite{shaffer23,passetti23}. Hence, as photons are injected into the cavity and trigger changes in the material, these modifications should become visible to measurements of the statistics of ejected photons. On generic grounds, materials near a quantum critical point (QCP) are a particularly attractive target to test these ideas, as strong quantum fluctuations should make them susceptible to light-induced changes. Light emitted from cavities hosting strongly-correlated and photon-dressed quantum materials should consequently exhibit strongly non-classical properties.

Here, we confirm these expectations, showing that the quantum statistics of light transmitted through cavities are direct evidence of quantum-electrodynamical changes to the state of matter in materials. Generalizing quantum optical input-output relations to strongly-correlated cavity quantum materials, we study the second-order photon coherence $g^{(2)}(t)$ for weakly-driven cavities hosting Mott insulators in the strong light-matter coupling regime. We find that transmitted photons exhibit pronounced (anti)bunching, which becomes drastically amplified near an Ising QCP, and establish measurements of $g^{(2)}$ as a probe of the ensuing photon-induced changes to the state of matter. We then show that quantum materials near criticality can in turn become collective single-photon sources, harnessing strong quantum fluctuations in the material to convert a laser drive into single photons. Finally, we demonstrate that chiral Mott insulators embedded in optical cavities can become efficient sources of Einstein-Podolsky-Rosen (EPR) pairs of entangled photons, as a resource for photon-based computation and sensing \cite{flamini2019}.

\section{Strong Light-Matter Coupling in Cavity Mott Insulators}

Strongly-correlated electron system with a large gap for charge excitations but Raman-active elementary excitations are a particularly attractive target for cavity-induced control. Paradigmatic examples are Mott insulators with partially filled valence orbitals, in which strong Coulomb repulsion localizes electrons and opens a large charge excitation gap. The low energy physics consist of magnetic interactions driven by virtual hoppings between nearby sites. When a two-dimensional Mott insulator is placed inside a cavity hosting a single infrared transverse electric (TE) mode with a frequency $\omega_0$ well below the material's charge gap, cavity photons cannot resonantly generate charge excitations. Instead, an effective interaction between photons and spins emerges from Raman processes, whereby a $q \approx 0$ photon is virtually absorbed and re-emitted into the cavity while exchanging pairs of spins as illustrated in Fig. 1(b). After integrating out charge excitations \cite{sentef20}, the photon-dressed spin exchange processes for spin-orbit-coupled $S=1/2$ cavity Mott insulators will generically take the form \seeAppendixX{\ref{app:cavityMott}}
\begin{align}
    \Ham_{\textrm{cav}} &= \sum_{ij,\alpha\beta} J_{ij}^{\alpha\beta}~ \hat{S}^\alpha_i \hat{S}^\beta_j ~+~ \aVecD \cdot \left( \omega_0 + \Dop \right) \cdot \aVec \label{eq:CavitySpinPhotonExchangeModel} \\
    \Dop &= \sum_{ij,\alpha\beta} \sum_{\mu\nu}  ~D_{ij,\mu\nu}^{\alpha\beta} ~\hat{S}^\alpha_i \hat{S}^\beta_j ~\otimes \polKetBra{\mu}{\nu},   \label{eq:Doperator}
\end{align}
consisting of Heisenberg, Ising, or off-diagonal couplings between spin-1/2 moments $\hat{S}_i^\alpha$ on atomic sites $i$ (with $\alpha = x,y,z$). The Hamiltonian in Eq. (\ref{eq:CavitySpinPhotonExchangeModel}) describes a static spin-exchange contribution $J_{ij}^{\alpha\beta}$, which includes the effect of vacuum fluctuations in a dark cavity, as well as a Fleury-Loudon \cite{fleuryloudon1968,shastryshraiman1990} Raman scattering term $D_{ij,\mu\nu} \propto g^2 (\mathbf{r}_{ij} \cdot \polVec_\mu ) ( \mathbf{r}_{ij} \cdot \polVec_\nu ) t^2/(U-\omega_0)$, which couples cavity photons $\aVec$ to spins. Here, $g$ is the light-matter coupling strength for electrons which scales as $\sim 1/\sqrt{\mathcal{V}}$ with the mode volume, $t$ describes electron hopping between sites, $U$ parameterizes the charge gap, $\mathbf{r}_{ij}$ is the vector connecting sites $i$ and $j$, $\aVec \equiv \sum_\mu \polVec_\mu \A{\mu}$ are the cavity photon operators, and $\polVec_\mu$ are the polarization vectors. This model is well-justified deep in the Mott phase $U\gg t$ for optical cavities $\omega_0 \gg J$. Here, multi-spin processes \cite{shastryshraiman1990} remain irrelevant, photon pair creation/annihilation processes $\sim (\AD{} \AD{} + \A{} \A{}) ~\hat{S}_i \hat{S}_j$ are energetically suppressed, and double-photon scattering $\sim \AD{} \AD{} \A{} \A{} ~\hat{S}_i \hat{S}_j$ remains negligible away from two-photon resonances and ultrastrong coupling \seeAppendixX{\ref{app:cavityMott}}.

In a dark cavity, vacuum fluctuations of the photon field weakly dress the material via $J_{ij}^{\alpha\beta}$. Similarly, as charge excitations are integrated out, ground state fluctuations in the material are weakly imprinted onto dark-cavity fluctuations of the photon field. However, these cannot be observed outside the cavity in the absence of an external drive \cite{ciuti06}.

Suppose instead that the cavity is weakly driven with photons. Since the Raman interactions conserve photon number, even a small number of photons in the cavity can have a profound impact on the material via repeated scattering. This requires that the cavity loss rate $\gamma$ is sufficiently small relative to the Raman spin-photon coupling strength, which scales as $D \sim \left[ e |\mathbf{r}_{ij}| / \sqrt{\hbar \omega_0 \epsilon \epsilon_0 \mathcal{V}} \right]^2 \times t^2/(U - \omega_0)$, where $e$ is the electron charge. For a 1000nm single-mode optical cavity with effective dielectric constant 4, a conservative estimate gives $D / J \sim 0.002 U / (U - \omega_0)$. This suggests that the requisite strong-coupling regime is accessible for Mott insulators in high finesse optical or infrared cavities by tuning the cavity mode $\omega_0$, rendering this class of systems a promising target for cQED control.

Consider now the perspective of photons tunneling through the cavity. If the material were to remain unperturbed by the cavity field, then transmitted photons would merely experience an effective linear dielectric medium. The bare cavity resonance $\omega_0$ would simply become renormalized to $\omega_c \equiv \omega_0 + \expect{ \polVecC_\mu \cdot \Dop \cdot \polVec_\mu }$ via the (mean field) ground state expectation value of the Fleury-Loudon operator for the polarization direction $\mu$. This would shift the resonance frequency for tunneling photons through the cavity, but leave their statistics unchanged.

On the other hand, any strong optical nonlinearity for photons transmitted through the cavity must necessarily arise as a backaction to photon-induced changes to the material's quantum state. After a photon enters and dresses the material due to strong light-matter coupling, the next photon arriving at the cavity will witness a modified system with shifted resonances. This effect is highly non-perturbative and relies on quantum fluctuations in the material as we will show. It not only changes the cavity transmission rate but also, crucially, modifies the photon statistics of light transmitted through the cavity. This immediately suggests that (1) a substantial modification of the photon statistics of transmitted light should serve as smoking-gun evidence of a cavity quantum-electrodynamical modification of the material, and (2) materials near a QCP, with large quantum fluctuations, are expected to be particularly susceptible to cavity-induced changes, making them an ideal target to observe this effect.

\section{Many-Body Input-Output Relations}

To quantify this effect, we now derive many-body input-output relations for the second-order photon coherence
\begin{align}
    g^{(2)}(t) = \frac{ \expectNormalOrder{ \Eminus(\tau) \Eminus(\tau+t) \Eplus(\tau+t) \Eplus(\tau) } }{ \expectNormalOrder{ \Eminus(\tau) \Eplus(\tau) } \expectNormalOrder{ \Eminus(\tau+t) \Eplus(\tau+t) } }  \label{eq:g2generic}
\end{align}
as a function of classical or quantum input light fields that are transmitted through a two-sided cavity hosting such materials. $g^{(2)}(t)$ describes the statistical correlation between pairs of emitted photons, and can be readily measured in a Hanbury Brown-Twiss setup [Fig. 1(a)]. $\mathcal{E}^{\pm}$ describe positive and negative frequency components of the electric field. Normal ordering is denoted by $:~:$ and is defined as $\Eminus$ acting to the left of $\Eplus$, with (anti) time ordering imposed on $\Eplus$ ($\Eminus$). For monochromatic input fields with frequency $\wIn$, which we will choose to be slightly detuned from the cavity frequency, $g^{(2)}$ depends only on the time delay $t$ between two photon detection events. Importantly, classical light fields must obey $g^{(2)}(0) \geq 1$, whereas antibunching of emitted photons with $g^{(2)}(0) < 1$ is a hallmark of quantum light \cite{scullyQuantumOptics}, which cannot arise from classical sources.

We start from a cavity coupled via tunneling amplitudes $\sqrt{\gamma/2\pi}$ to ``left'' (L) and ``right'' (R) free-space photon bath modes $\bVecX{L/R}(\omega)$, which are driven by an external input field [Fig. 1(a)]. The input states for side $\alpha = L,R$ are defined as the Fourier transform of free-space mode frequencies $\bInVecX{\alpha}(t) = (1/\sqrt{2\pi}) \int d\omega e^{-i\omega(t-t_\mathrm{in})} \bVecX{\alpha}(\omega; t_\mathrm{in})$ at time $t_\mathrm{in} \to -\infty$. Observed fields at the detector $\Eplus$ can be replaced via quantized output fields $\bOutVecX{R}$, which are defined analogously as $\bOutVecX{\alpha}(t) = (1/\sqrt{2\pi}) \int d\omega e^{-i\omega(t-t_\mathrm{out})} \bVecX{\alpha}(\omega; t_\mathrm{out})$ with $t_\mathrm{out} \to +\infty$ \cite{collett1984squeezing,gardiner85}. A Markov approximation with unbounded free-space mode frequencies and frequency-independent photon tunneling is justified by the large disparity between the bare cavity frequency $\omega_0$ and elementary excitations in the material. Input and output fields for the same side $\alpha$ then satisfy the input-output relation $\bOutVecX{\alpha}(t) = \bInVecX{\alpha}(t) - i\sqrt{\gamma} \aVec(t)$ as a function of the cavity photon operator $\aVec(t)$ in the Heisenberg picture \cite{gardiner85}.

\begin{figure*}[t!]
    \centering
    \includegraphics[width=\textwidth]{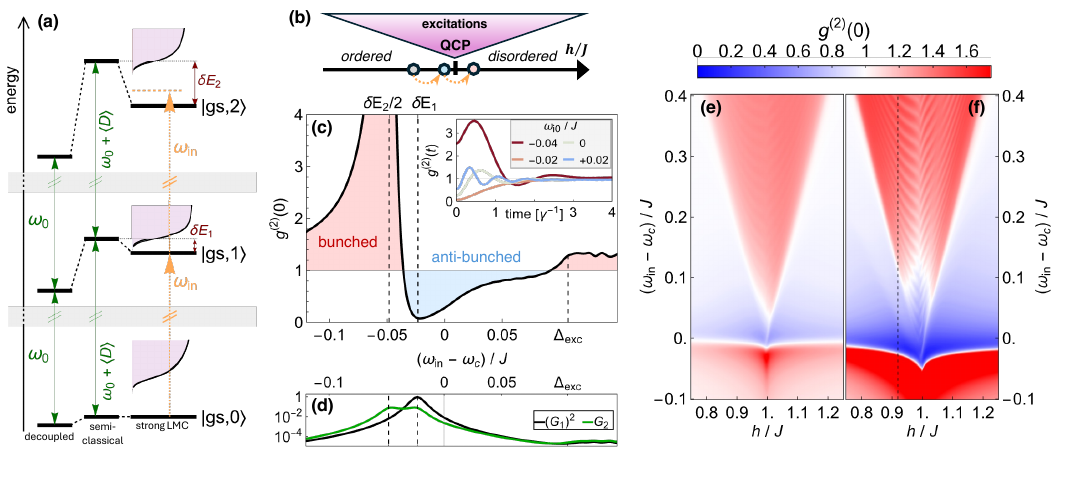}
    \caption{\captiontitle{Photon Antibunching Near a Quantum Critical Point.} (a) Photon blockade in quantum many-body systems, depicted for an almost-critical 1D Ising chain coupled to cavity photons. Starting from (left) a bare cavity mode with frequency $\omega_0$, (middle) inserting a material induces a small dark-cavity shift and renormalizes the cavity frequency to $\omega_c = \omega_0 + \expect{\hat{D}}$. For strong light-matter coupling (LMC), (right) the $n$-photon-dressed ground state $\ket{{\rm gs},n}$ and excitations are nudged across a QCP (b), generating a large single-photon non-linearity from quantum fluctuations in the material. This results in bunching and antibunching of transmitted photons as a function of input photon detuning, depicted in (c) for $h/J = 0.92$ [dotted line in (f)]. Vertical dashed lines denote single- and two-photon resonances $\delta E_1$ and $\delta E_2 / 2$, as well as the material excitation gap. Inset shows the second-order photon coherence $g^{(2)}(t > 0)$ at finite time delay upon sweeping the input photon frequency. (d) Single- and two-photon tunneling rates are peaked near the single- and two-photon resonances. (e), (f) $g^{(2)}(0)$ shown as a function of the input photon detuning and transverse field for (e) weak [$D/J = 0.0065, \gamma/J = 0.015$] and (f) strong [$D/J = 0.0125, \gamma/J = 0.01$] LMC. }
    \label{fig:ising_coincidence}
\end{figure*}

To connect $\bOutVecX{R}(t)$ to $\bInVecX{L}(t)$, one must solve a many-body Langevin equation for the cavity quantum material, establishing an relation between input and output correlation functions. Vacuum expectation values of $\aVec(t)$ vanish in the absence of a drive due to dissipation. After integrating out the bath modes \seeAppendixX{\ref{app:inputOutputMott}}, the cavity photon operators obey the equations of motion
\begin{align}
    i\frac{\partial}{\partial t} \aVec(t) &= \left[ \omega_0 - i\gamma + \Dop(t) \right] \cdot \aVec(t) + \sqrt{\gamma}~ \bInVec(t),  \label{eq:cavityEoM}
\end{align}
where $\bInVec \equiv \bInVecX{L} + \bInVecX{R}$. Formally integrating Eq. (\ref{eq:cavityEoM}) and substituting into the input-output relation yields
\begin{align}
    \bOutVecX{R}(t) = \bInVecX{R}(t) - \gamma \hspace{-0.05cm}\int\limits_{-\infty}^t \hspace{-0.05cm} d\tau e^{i(\omega_0-i\gamma)(\tau-t)} \Uop(t,\tau) \cdot \bInVec(\tau)  \label{eq:inputOutput}
\end{align}
where 
\begin{align}
    \Uop(t,t') = \TimeOrder \exp\left\{ -i \int_{t'}^{t} d\tau ~\Dop(\tau) \right\} \label{eq:scatteringOperator}
\end{align}
is a time-ordered propagator that obeys ${\partial_t \Uop(t,t') = -i \Dop(t) \cdot \Uop(t,t')}$ and describes the repeated polarization-dependent scattering of photons off the material for as long as they reside within the cavity. Causality dictates that cavity material operators $\Dop(t')$ cannot depend on the input field $\bInVec(t)$ at a later time $t > t'$; such operator pairs thus commute $\left[ \BinX{\alpha\mu}(t), \hat{D}_{\mu\nu}(t') \right] = 0$, however the same is not true for $t < t'$ \cite{gardiner85}. Eq. (\ref{eq:g2generic}) must therefore be carefully evaluated \cite{xu2015input} by time-ordering while respecting the causality relations \seeAppendixX{\ref{app:inputOutputG2}}.

In transmission geometry, the input field arrives from the left side, and the right side input is in a vacuum state [$\bInVecX{R} \ket{\rm in} = 0$].
At a high level, the Langevin equation then relates $n$-point correlation functions of output photons at a detector to $n$-point input correlators
\begin{align}
    &\expectNormalOrder{ \bOutVecXD{R}(t_1) \cdots \bOutVecXD{R}(t_n) \bOutVecX{R}(t'_n) \cdots \bOutVecX{R}(t'_1)}  \notag\\
        &\hspace{0.2cm}\sim \int d\tau_1 d\tau'_1 \cdots ~\mathbf{K}^{(n)}_{\rm mat}(t_1, t'_1 \cdots t_n, t'_n;\tau_1,\tau'_1\cdots\tau_n,\tau'_n) \notag\\
        &\hspace{0.5cm} \cdot \expectNormalOrder{ \bInVecXD{L}(\tau_1) \cdots \bInVecXD{L}(\tau_n) \bInVecX{L}(\tau'_n) \cdots \bInVecX{L}(\tau'_1) },
\end{align}
where $\mathbf{K}^{(n)}_{\rm mat}$ is a highly non-perturbative function of the material that arises from expanding $\Uop$ and formally describes an infinite series of Raman scattering diagrams $\propto \expect{ \cdots \Dop(\tau_i) \Dop(\tau_j) \cdots }$ for $n$ photons. We compute this explicitly for $g^{(2)}(t)$ below.

In principle, the Heisenberg-picture spin-photon operators $\Dop(t)$ have an implicit dependence on the input fields. Made explicit, these terms encode subleading dependencies on higher-order $m > n$ input-photon correlations, which reflect the possibility of additional photons entering the cavity before the first $n$ photons are emitted. Such events, however, can be neglected for weak input fields and high quality cavities, which amounts to taking $\Dop(t) \approx e^{-i\Ham t} \Dop e^{i\Ham t}$, where $\Ham$ is the material Hamiltonian. Stronger drive strengths can be handled using the Keldysh formalism \cite{sieberer2016keldysh,talkington2024linear,talkington2025}.

The physical relation between input and output fields becomes particularly simple in quasi-one-dimensional systems, which only couple to cavity photons linearly polarized along the active material axis (denoted $x$), such that the dependence on photon polarization can be neglected. Taking $\BinX{L}\equiv\bInVecX{L}\cdot \polVec_x$ and $\hat{D} \equiv \polVecC_x \cdot \Dop \cdot \polVec_x$ in Eq. (\ref{eq:inputOutput}) and substituting into Eq. (\ref{eq:g2generic}),
$g^{(2)}$ for $x$-polarized photons can be computed by carefully evaluating the time- and normal-ordered products \seeAppendixX{\ref{app:inputOutputG2}}. At coincidence, we find
\begin{align}
    g^{(2)}(0) &= \frac{ \sum\limits_f \left| \bra{f} \frac{2}{2\wIO + E_0 - \Ham - 2\hat{D} + 2i\gamma} \frac{1}{\wIO + E_0 - \Ham - \hat{D} + i\gamma} \ket{0} \right|^2}{ \left[ \sum\limits_f \left| \bra{f} \frac{1}{\wIO + E_0 - \Ham - \hat{D} + i\gamma} \ket{0} \right|^2 \right]^2} \notag\\
    &\times  \frac{ \expect{ \BinXD{L} \BinXD{L} \BinX{L} \BinX{L} }(\wIn) }{ \left[ \expect{ \BinXD{L} \BinX{L} }(\wIn) \right]^2 }, \label{eq:g2_1D}
\end{align}
where $\ket{0}$ is the material ground state with energy $\Ham \ket{0} = E_0 \ket{0}$, $\ket{f}$ denotes a complete set of final states for the material after a pair of photons leave for the detector, and $\wIO \equiv \wIn - \omega_0$ is the input field detuning from the bare cavity resonance. The second line describes the coherence of the input field, and is equal to one for a classical coherent state.

Eq. (\ref{eq:g2_1D}) is a central theoretical result of this work and has an appealing physical interpretation. Formally, it describes the tunneling of pairs of photons through the cavity, normalized by the squared single-photon tunneling rate. The numerator models a simple sequential process. Starting from the material ground state, a photon tunnels into the cavity, and the material suddenly becomes subjected to a photo-modified Hamiltonian $\Ham + \hat{D}$. This quench can be viewed as an infinite-order Raman scattering process via expanding the resolvent $1 / (\wIO + E_0 - \Ham - \hat{D})$ in powers of $\hat{D}$. Subsequently, a second photon tunnels into the cavity with an already modified material, further changing the cavity Hamiltonian to $\Ham + 2 \hat{D}$. Finally, both photons tunnel out toward the detector, leaving the material in a final state $\ket{f}$. A factor of two accounts for permutations of the observed photon pair. The denominator describes an analogous process for tunneling each photon through the cavity independently. A derivation of $g^{(2)}(t)$ at finite time delay $t>0$ is similar, though must account for additional processes whereby the first photon is ejected before the second enters the cavity \seeAppendixX{\ref{app:inputOutputG2}}.

\section{Many-Body Photon Blockade}

We now relate the propensity for bunching or antibunching of transmitted photons to cavity-induced modifications of the material near a QCP. To illustrate this, we study the transverse-field Ising model (TFIM), realized in a paradigmatic class of candidate systems such as CoNb\textsubscript{2}O\textsubscript{6} \cite{maartense1977,coldea2010,woodland2023}. External magnetic fields can tune such materials across an Ising QCP separating ferromagnetic from paramagnetic phases. In its simplest form, the Hamiltonian for spins coupled to cavity photons takes the form $\Ham_{\rm cav} = \Ham + (\omega_0 + \hat{D}) \AD{} \A{}$ with
\begin{align}
    \Ham &= -2J \sum_i \hat{S}_i^z \hat{S}_{i+1}^z + h \sum_i \hat{S}_i^x~\\
    \hat{D} &= 2D \sum_i \hat{S}_i^z \hat{S}_{i+1}^z.
\end{align}
Here, $J$ and $h$ denote nearest-neighbor FM Ising exchange and the transverse field.
For simplicity, we ignore subleading Kitaev and $\Gamma$ exchange contributions found in actual materials \cite{konieczna2024,churchill2024}, and absorb small dark-cavity spin-exchange corrections due to vacuum fluctuations into the definition of $J$. The QCP lies at $h=J$.

Fig. \ref{fig:ising_coincidence}(a) depicts the hierarchy of eigenstates of a cavity Ising magnet dressed by $n$ photons near criticality, with parameters chosen such that the two-photon state crosses the QCP [Fig. \ref{fig:ising_coincidence}(b)]. The energy of the $n$-photon-dressed material ground state $\ket{{\rm gs},n}$ is $E_n = E_0 + n \expect{\hat{D}}_0 + \delta E_n$, where $E_0$ is the dark-cavity ground state energy, $n \expect{\hat{D}}_0$ (evaluated for the dark cavity ground state) describes the mean-field cavity resonance shift depicted in the middle column of (a), and $\delta E_n$ is the nonlinearity arising from quantum fluctuations in the material depicted in the right column. To leading order in $D$, the nonlinearity scales as $\delta E_n \approx n^2 \Real\{ \mathcal{C}(0) \}$ where $\mathcal{C}(\omega) = -i \int_0^\infty dt ~ e^{i\omega t} \expect{ \delta\hat{D}(t) \delta\hat{D}(0) }_0$ is the material Green's function for fluctuations $\delta\hat{D} \equiv \hat{D} - \expect{\hat{D}}$. In the thermodynamic limit, the fluctuations diverge at the QCP, which however is offset by the inverse-mode-volume scaling of $D$. Nevertheless, for finite cavities with optimally-compressed mode volumes and fixed light-matter coupling $D$, this means that the material-induced nonlinearity becomes dramatically enhanced at the QCP.

Due to the opposing signs of Ising ($J$) and Raman ($D$) exchange couplings, the true $n$-photon cavity resonances lie below the mean field value $\omega_c$ by $\delta E_n\propto -n^2$. This implies that if the input field is detuned from the cavity frequency ($\wIC \equiv \wIn - \omega_c$) such that it is one-photon resonant ($\wIC = \delta E_1$), it will necessarily be blue-detuned from the two-photon resonance ($\wIC = \delta E_2 / 2$) and vice versa. So long as the disparity between one- and two-photon resonances exceeds the cavity decay rate ($|\delta E_1-\delta E_2/2 | > \gamma$), the cavity quantum material will produce a pronounced \textit{photon blockade effect} at the one- and two-photon resonance frequencies. This is illustrated in Fig. \ref{fig:ising_coincidence}(c), where input frequencies near $\omega_c+\delta E_1$ exhibit significant antibunching as single photon transmission becomes resonantly enhanced and two photon transmission is blocked, while the reverse occurs for frequencies near $\omega_c+\delta E_2/2$, where photon bunching is observed.

Fig. \ref{fig:ising_coincidence}(e) and (f) depicts $g^{(2)}(0)$ for the cavity TFIM as a function of transverse field, calculated for weak- and strongly-coupled cavities respectively \seeAppendixX{\ref{app:TFIMsims}}.  A transition from bunching to antibunching as a function of input detuning emerges as a generic feature in $g^{(2)}(0)$ probed near the cavity resonance $\wIC\approx0$. This blockade effect becomes drastically enhanced for high-fidelity cavities near the QCP [Fig. \ref{fig:ising_coincidence}(f)]. At higher input frequencies, $g^{(2)}(0)$ probes the magnetic excitation spectrum; as $\wIn$ becomes two-photon resonant with excitations above the two-photon ground state, the statistics of transmitted light reverts to bunching. Therefore, the ``fan'' of magnetic excitations above the QCP is reflected in enhanced bunching of transmitted light for blue-detuned inputs.

\section{Witnessing Light-Induced Material Changes}

Experimentally, this connection between cavity-induced energy shifts and photon blockade can be inverted, with photon coincidence measurements acting as a witness for cavity quantum-electrodynamical material changes. In particular, measuring $g^{(2)}(t \to 0)$ as a function of input frequency and extracting the detunings $\wIC$ at which maximal antibunching and bunching occur provides an estimate for the photo-induced energy shifts of the magnetic state, $\delta E_1$ and $\delta E_2$ respectively, as a proxy of photo-induced material changes. Importantly, photon coincidence measurements can distinguish light-matter coupling from other, spurious changes to the material/cavity. For example, while changes to the dielectric screening environment or charge accumulation at the material-mirror interface would be expected to shift the cavity frequency $\omega_c$, they would not affect nor be reflected in photon statistics.

\begin{figure*}[!ht]
    \centering
    \includegraphics[width=\textwidth]{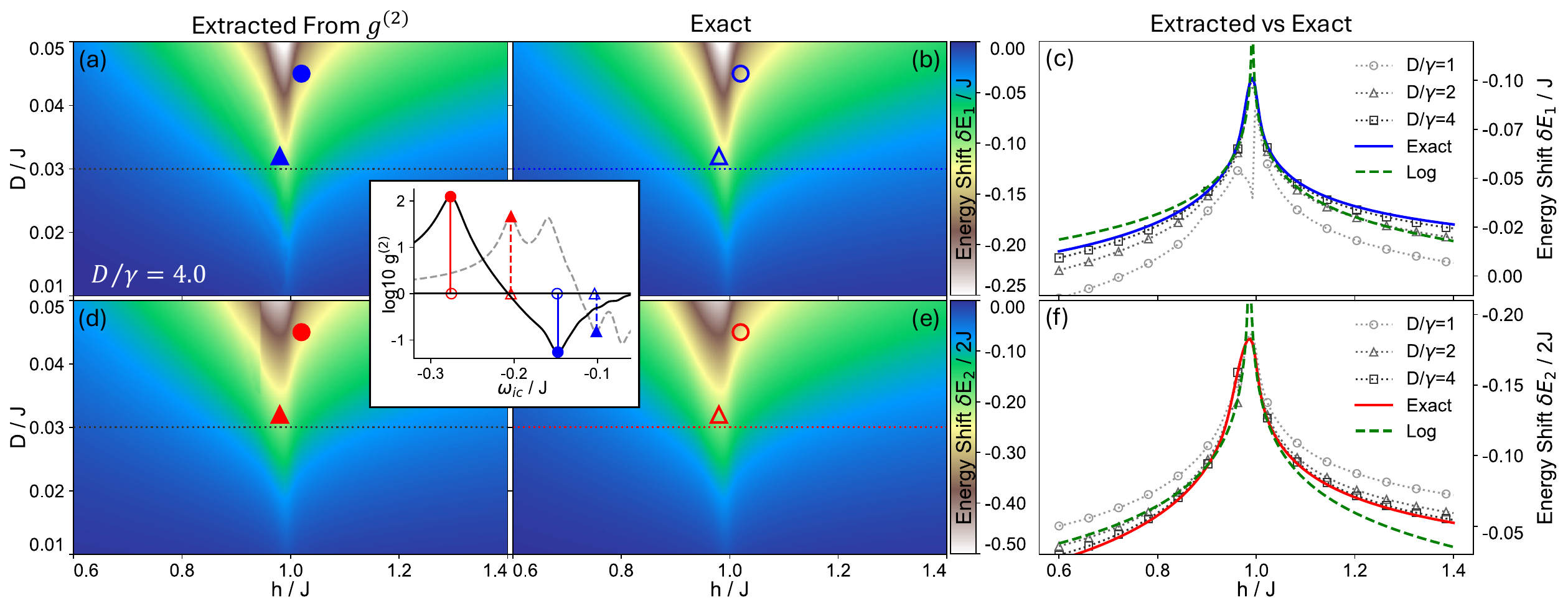}
    \caption{\captiontitle{Probing Cavity-Induced Material Changes via Photon Statistics.} (a) The input frequency detuning $\wIC/J$ that gives maximal antibunching at $D/\gamma=4.0$ provides an accurate measurement of (b) the single photon induced energy shift $\delta E_1/J$ calculated for a 256 site TFIM chain. (c) This estimate converges as $\gamma\to0$ and has logarithmic critical scaling at the QCP, shown for the dotted lines across (a) and (b) at $D/J=0.3$. (d), (e), and (f) Similarly, frequencies $\wIC/J$ giving maximal bunching measure half the two photon induced energy shift $\delta E_2/2J$. Inset: two examples comparing the maximal antibunching peaks (solid blue) vs $\delta E_1/J$ (open blue) and the maximal bunching peaks (solid red) vs $\delta E_2/2J$ (open red), corresponding to the matched symbols in (a), (b), (d), and (e). }
    \label{fig:ising_energy_shifts}
\end{figure*}

Figure \ref{fig:ising_energy_shifts} (a) depicts the optimal detunings $\wIC$ for which transmitted photons become maximally antibunched while (b) shows $\delta E_1$ computed directly for the TFIM \seeAppendixX{\ref{app:TFIMsims}}. Similar results for maximal bunching $\wIC$ and $\delta E_2 / 2$ are shown in (d) and (e) respectively. One immediately sees two striking features. First, by comparing the panels, it is clear that photon coincidence measurements can faithfully monitor cavity-induced material changes. The measurement fidelity improves with the quality of the cavity and proximity to the QCP [Figs. \ref{fig:ising_energy_shifts}(c) and (f)]. On the other hand, if $|\delta E_1-\delta E_2/2 | \lesssim \gamma$, then the bunching and antibunching peaks can no longer be clearly resolved due to broadening of the one- and two-photon resonances. In such cases, $\wIC$ for maximal antibunching is blueshifted from the $\delta E_1$ resonance [(c)] while the maximal bunching peak is redshifted [(f)], leading to a slight over- (under-) estimation of $\delta E_2$ ($\delta E_1$).

Second, the observable enhancement of these photon-induced changes near a QCP is a direct consequence of the critical scaling behavior around $h/J = 1$. Formally, for small $D/J$, and as a correction to the mean-field contribution $\sim n \expect{\hat{D}}$ [Fig. \ref{fig:ising_coincidence}(a)], the $n$-photon-induced energy shift $\delta E_{n}$ of the material \textit{within the mode volume} $\mathcal{V}$ diverges logarithmically as
\begin{align}
    \delta E_{n} \sim n^2 \mathcal{V} \frac{D^2}{J} \log \left| \frac{h - J}{8J/e^2} \right|.
\end{align}
This observable scaling behavior is depicted in Fig. \ref{fig:ising_energy_shifts}(c) and (f). A small apparent shift in $h/J$ of the critical point in (c) and (f) reflects the movement of the QCP as one or two photons tunnel into the cavity. The formal apparent divergence stems from the non-analyticity of the QCP and, importantly, is cured by finite-size effects, since the many-body energy shift of the cavity quantum material derives solely from spins within the optical mode volume $\mathcal{V}$. For typical optical cavities, however, this encompasses thousands of spins, which now act in concert as a single highly-correlated emitter near the QCP.

\section{Cavity Quantum Materials as Single-Photon Sources}

An enticing consequence of quantum criticality enhanced photon blockade is the prospect for constructing an efficient single-photon source from cavity-embedded quantum materials. 
Single emitter photon blockade is a well-established approach for single-photon generation, but has efficiency limited by the emitter's cross section relative to the cavity mode area \cite{Goto2019,Reiserer2015}. Using multiple emitters improves light-matter coupling, but reduces the single-photon nonlinearity for uncorrelated atoms \cite{Trivedi2019}, requiring schemes to induce cooperativity \cite{Chen2022}. The inherent strong correlations of quantum materials near the QCP provides a direct means to harness a macroscopic system of $N$ spins acting together like a single extended emitter. 

\begin{figure*}[!htb]
    \centering
    \includegraphics[width=\textwidth]{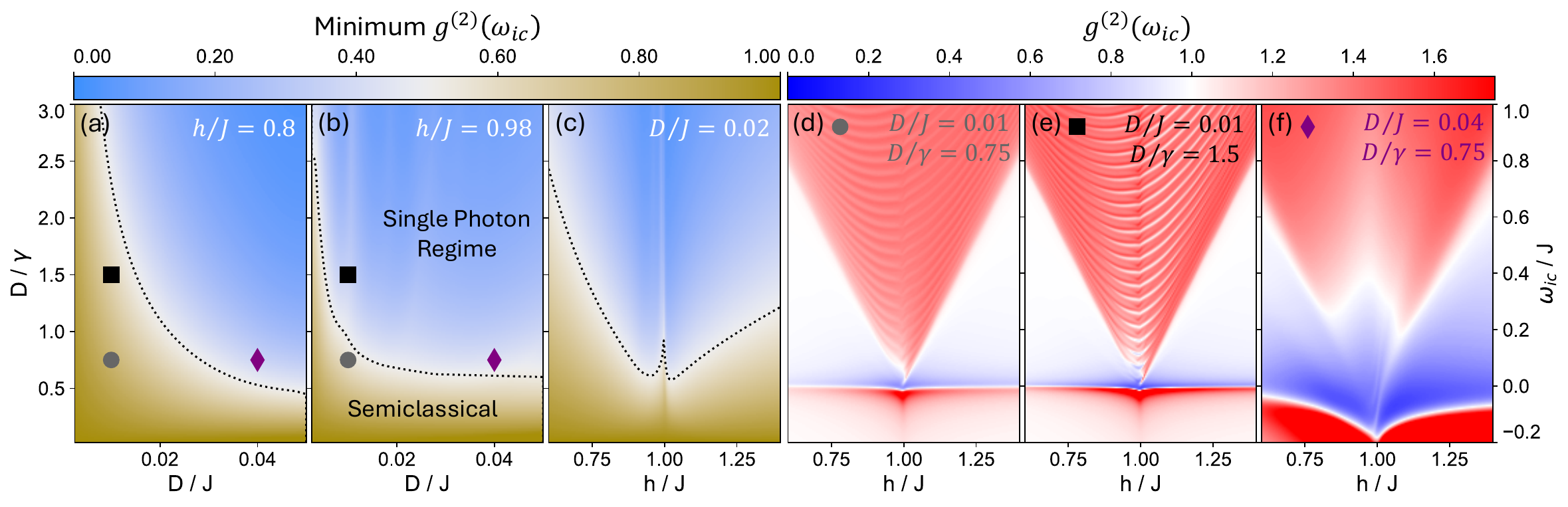}
    \caption{\captiontitle{Generation of Single Photons from Cavity Quantum Materials.} Cuts through the parameter space showing the maximum value of antibunching for a 256 site TFIM chain at $h/J=0.8$ (a), $h/J=0.98$ (b), and $D/J=0.02$ (c). The black dotted lines represent the crossover to regions where $g^{(2)}(0)$ can be less than 0.5 and the cavity-embedded material acts as a single collective emitter. (a) and (b) highlight that quantum light generation is enhanced by increasing light-matter coupling and improving the cavity quality. (c) shows how proximity to the QCP enhances the collective single-photon nonlinearity. The full coincidence spectra are shown for three values of $D/J$ and $D/\gamma$ corresponding to the gray (d), purple (e), and green (f) points indicated, showing the enhancement of antibunching with light-matter coupling and cavity strength, in proximity to the QCP.}
    \label{fig:ising_antibunching}
\end{figure*}

To assess the efficacy of quantum materials as single-photon sources, Fig. \ref{fig:ising_antibunching} shows the maximum value of antibunching obtained as a function of light-matter coupling strength and cavity quality [(a) and (b)], as well as proximity to the QCP [(c)]. Regions where the system acts as a single extended emitter ($g^{(2)}<0.5$) are shown in blue, with a black dotted line delineating the first point where $g^{(2)}=0.5$. The results are very similar to what was found for light-induced energy shifts. Figures (a) and (b) highlight that optimal antibunching increases with light-matter coupling strength and cavity quality. For the most part, antibunching also increases as one approaches the QCP, which might be expected as increasing correlations in the material will cause it to behave more like a single extended emitter. This trend reverses, however, as one gets too close to the QCP.

To understand this better, Figs. \ref{fig:ising_antibunching}(d), (e), and (f) show photon coincidence as a function of $\wIC$ and $h$ for the points indicated in (a) and (b). Fig. \ref{fig:ising_antibunching}(d) depicts light-matter coupling strength $D$ in the perturbative regime, where photon statistics probe the material excitation spectrum. (e) shows the enhancement of both bunching and antibunching as the cavity quality is increased while holding $D$ fixed. Individual excitations of the finite size system appear better resolved as their linewidths decrease with $\gamma$, but their energies remain unchanged. By contrast, the behavior as we increase the light-matter coupling strength in (f) is dramatically different. In addition to increasing the strength of the response, the one- and two-photon ground state resonances move to lower frequencies as $\delta E_1$ and $\delta E_2$ increase. At the same time, two-photon resonances with excited states can overlap with the single-photon resonance [inset to Fig. \ref{fig:ising_energy_shifts}], with offsetting effects that can reduce the maximal antibunching found in the vicinity of the QCP.

\section{Cavity Quantum Materials as Entangled EPR Pair Sources}

\begin{figure}
    \centering
    \includegraphics[width=1.05\columnwidth]{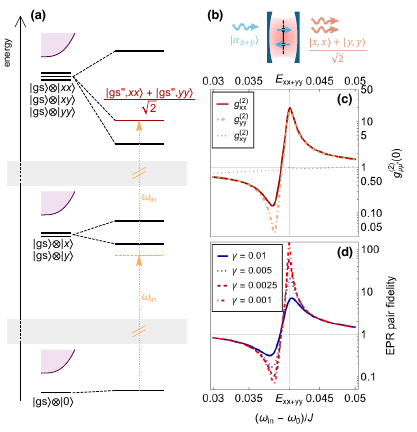}
    \caption{\captiontitle{Generation of EPR Pairs from Cavity-Embedded Chiral Materials.} (a) Schematics of the polarization-selective many-body photon blockade, depicting zero-, one-, and two-photon-dressed material ground states and excitations. In chiral quantum materials, the two-photon-dressed ground state splits into an EPR-pair-dressed state $\ket{\textrm{gs}''}$ and a pair of chiral states. (b) If a classical $(x + y)$ polarized input field is two-photon resonant with the EPR-pair-dressed ground state [(a)], then transmitted photons are ejected as bunched EPR pairs, constituting a source for entangled photons. (c) Signatures of EPR pair generation in polarization-sensitive measurements of $g_{\mu\mu'}^{(2)}(t=0)$ depicted for a chiral magnetic material, showing strong bunching of pairs of $x$ and $y$ polarized photons while suppressing $xy$ photon pair transmission. (d) Fidelity of EPR pair generation, depicted via the ratio of $g_{xx}^{(2)}(0) / g_{xy}^{(2)}(0)$ as a function of cavity loss rates $\gamma$ (in units of exchange $J$).}
    \label{fig:EPR}
\end{figure}

Armed with cavity quantum materials near a QCP as sources of single photons, it is intriguing to instead consider input or output fields with \textit{quantum-entangled} polarization components and inquire whether photon entanglement can be either extracted from or transferred onto the material. We therefore study the second-order photon coherence $g_{\mu\mu'}^{(2)}(t)$ for detected output photon pairs with polarization $\polVec_\mu, \polVec_{\mu'}$. Intuitively, the transmission of polarization-entangled input states can be understood as a sum over trajectories with semi-classical inputs. Conversely, even a classical input field can generate polarization-entangled outputs.

Formally, the one- and two-photon input correlation functions can be decomposed
\begin{align}
    \expect{ \BD{\mathstrut \nu} \B{\mathstrut \nu'} }_{\textrm{in}} &\equiv \sum_j \left[ f_{\mathstrut \nu}^{(j)} \right]^* f_{\mathstrut \nu'}^{(j)} \\
    \expect{ \BD{\mathstrut \nu_1} \BD{\mathstrut \nu'_1} \B{\mathstrut \nu'_2} \B{\mathstrut \nu_2} }_{\textrm{in}} &\equiv \sum_{j} \left[ \sum_i  F^{(ij)}_{\mathstrut \nu_1} F^{(ij)}_{\mathstrut \nu'_1} \right]^* \left[ \sum_{i'} F^{(i'j)}_{\mathstrut \nu'_2} F^{(i'j)}_{\mathstrut \nu_2} \right]
\end{align}
into sums of semi-classical one- and two-photon inputs with amplitude-weighted polarization vectors $\fVec{j}$ and $\FVecIJ{i}{j}$, where $j$ indexes the trajectory and $i$ is an internal index. A coherent input state with polarization $\polVec_{\rm in}$ and amplitude $\alpha_{\rm in}$ is described by a single trajectory $\fVec{0} = \FVecIJ{0}{0} = \alpha_{\rm in}~ \polVec_{\rm in}$. Similarly, $n$-photon number states with fixed polarization are described by $\fVec{0} = \sqrt{n} \polVec_{\rm in}$, $\FVecIJ{0}{0} = [n(n-1)]^{1/4} \polVec_{\rm in}$. Conversely, consider the EPR pair
\begin{align}
    \ket{\Psi} = \frac{ \BinXD{\updownarrow} \BinXD{\updownarrow} + \BinXD{\leftrightarrow} \BinXD{\leftrightarrow} }{\sqrt{2}} \ket{0}  \label{eq:bellState}
\end{align}
with frequency $\wIn$ and vertical (horizontal) polarization $\updownarrow$ ($\leftrightarrow$) in the material plane. Here, at least two trajectories are needed to construct the input-output relation, describing the transmission of a $\updownarrow$-polarized ($\fVec{0} = \FVecIJ{0}{0} = \polVec_{\updownarrow}$) and a $\leftrightarrow$-polarized ($\fVec{0} = \FVecIJ{0}{0} = \polVec_{\leftrightarrow}$) photon pair. Applying the non-perturbative input-output relation [Eq. (\ref{eq:inputOutput})], the polarization-resolved output photon coincidence $g_{\mu\mu'}^{(2)}(0)$ generalizes Eq. (\ref{eq:g2_1D}) and readily follows as \seeAppendixX{\ref{app:inputOutputG2}}
\begin{align}
    g_{\nu\nu'}^{(2)}(0) = \frac{ \sum\limits_{j,f} \left| \sum\limits_i \bra{f} \left[ \polTensorC_{\nu\nu'} \hat{\boldsymbol{G}}_2 \FVecIJ{i}{j} \otimes \left[ \hat{\boldsymbol{G}}_1 \FVecIJ{i}{j} \right] \right] \ket{0} \right|^2  }{  \prod\limits_{\sigma=\nu,\nu'} \bigg(\sum\limits_{j,f} \left| \bra{f} \polVecC_\sigma \hat{\boldsymbol{G}}_1 \fVec{j} \ket{0} \right|^2\bigg)  }.  \label{eq:g2withpolarization}
\end{align}
Here, $\polTensor_{\nu\nu'} = \polVec_\nu \otimes \polVec_{\nu'} + \polVec_{\nu'} \otimes \polVec_{\nu}$ is symmetrized two-photon polarization vector for output polarizations $\nu,\nu'$. The dressed many-body Green's resolvents for the cavity material
\begin{align}
    \hat{\boldsymbol{G}}_1 &= \frac{1}{\wIO + E_0 - \Ham - \Dop + i\gamma} \\
    \hat{\boldsymbol{G}}_2 &= \frac{1}{2\wIO + E_0 - \Ham - \Dop \otimes 1 - 1 \otimes \Dop + 2i\gamma}
\end{align}
describe the change upon addition of 1 ($\hat{\boldsymbol{G}}_1$) or 2 ($\hat{\boldsymbol{G}}_2$) photons, and consequently act on a product state of material excitations  and the two-dimensional (four-dimensional) space of polarizations for 1 (2) photons. The indistinguishability of two-photon $\ket{xy}$ and $\ket{yx}$ states (in a basis of linear polarizations) is enforced as a symmetry of Eq. (\ref{eq:g2withpolarization}) \seeAppendixX{\ref{app:inputOutputG2}}.

For polarization-entangled input states, the detected output field simply accesses an interferometric superposition of distinct light-matter-dressed states in the cavity. For instance, driving cavities hosting a square-lattice antiferromagnet such as La$_2$CuO$_4$ with linearly polarized EPR state input [Eq. (\ref{eq:bellState})] can reveal cavity-photon-induced renormalizations of \textit{both} $x$ and $y$ bond magnetic exchange couplings (via $\hat{D}_{xx}$, $\hat{D}_{yy}$ in a one-shot measurement.

Strikingly however, strongly light-matter-coupled 2D cavity materials can \textit{generate} entangled output photon pairs from classical inputs. 
The necessary ingredients can be inferred from symmetry alone. Suppose that a 2D Mott insulator with a gapped magnetic ground state is placed inside an optical cavity. If the material has four-fold rotation symmetry (which ensures $C_4 \hat{D}_{xx} C_4^\dag = \hat{D}_{yy}$, $C_4 \hat{D}_{xy} C_4^\dag = -\hat{D}_{yx}$), then the two-photon tunneling rates for a pair of $x$- or $y$-polarized photons through the cavity must be identical. However, the tunneling rate for a pair of one $x$- and one $y$-polarized photon can differ, provided that no mirror symmetries are present (mirror would enforce $M \hat{D}_{xy} M^\dag = -\hat{D}_{xy}$). 

Fig. \ref{fig:EPR}(a) schematically depicts the resulting polarization-selective many-body photon blockade for such chiral materials. One-photon dressed matter states split into chiral $\ket{gs',x\pm iy}$ pairs. Importantly however, the two-photon dressed manifold splits into chiral combinations as well as a matter state dressed by an EPR pair [Eq. (\ref{eq:bellState})]. As the material ground state transforms trivially under all point group symmetries, such a splitting of two-photon-dressed states necessitates that all mirror symmetries are broken. Now, if a classical input field is two-photon-resonant with this state, and one-photon resonances are again avoided (due to matter fluctuations), then one immediately finds that light emitted from the cavity must come primarily in bunches of EPR pair states [Fig. \ref{fig:EPR}(b)], while single-photon emission and emission of $x,y$ photon pairs is suppressed. Fig. \ref{fig:EPR}(c) illustrates the resulting signatures in polarization-resolved coincidence counting measurements of $g_{\mu\mu'}^{(2)}(t=0)$, computed for a minimal toy model of a cavity-embedded chiral quantum magnet on a four-site plaquette \seeAppendixX{\ref{app:EPRsims}}.
In analogy to single-photon generation, achieving high fidelity EPR pair generation again requires that photon-induced energy shifts are resolvable within the cavity linewidth [Fig. \ref{fig:EPR}(d)], and is greatly augmented by strong quantum fluctuations near a critical point. Enticingly, this suggests that strongly-correlated chiral cavity quantum materials can serve as a novel source of quantum-entangled photon states, suitable for photon-based quantum computation.

\section{Outlook}

This paper shows that the strong light-matter coupling regime of cavity-embedded quantum materials can lead to profound changes to the statistics of transmitted light, particularly when tuned near a QCP. This serves as smoking gun evidence for light-induced changes to the state of matter, while simultaneously establishing quantum-critical materials as a novel source for quantum light generation.

Immediate applications of these results are two-fold. First, measurements of photon (anti)bunching can provide a broader tool to resolve competing instabilities in strongly-correlated quantum materials, by selectively dressing their elementary excitations with light and again inferring the resulting change to the state of matter from the statistics of emitted photons. For instance, polarization-selective measurements of $g^{(2)}(t)$ could provide a cQED route to shed light on the pairing glue in unconventional superconductors, via placing them inside multi-mode cavities designed to resonantly dress collective modes such as phonons, amplitude, or density oscillations. Generalizations to spatiotemporal multi-point measurements of higher-order photon coherences are a particularly intriguing direction for accessing highly-entangled states in cavity quantum materials. Here, correlating the detection of photons ejected from spatially-distinct regions of a cavity quantum material (demarcated via the cavity design) can be used to probe the spread of non-local material correlations, with intriguing connections to entanglement witnesses or topological order. Finally, a very recent work has shown that Hanbury-Brown-Twiss measurements of the two-photon scattering off materials can harbor signatures of non-local correlations and fractional statistics of emergent anyonic excitations \cite{nambiar2024diagnosing}. It will be interesting to generalize these ideas to cavities in the strong light-matter coupling regime, to understand non-perturbative cavity-induced modifications of proximal quantum spin liquid materials from the quantum statistics of transmitted photons.

Second, the use of cavity quantum materials as sources of single photons or EPR pairs can enable new functionalities for photon-based quantum computation. Electrical or magnetic control of a material near a critical point can tune the relation between input and output light on demand---serving in principle as a switchable few-photon nonlinearity. Furthermore, stronger drives and highly-entangled quantum states of matter could be harnessed to produce $n$-photon resource states such as GHZ and W states that can serve as entanglement sources for computation \cite{walther2005experimental}. Theoretically, strong illumination of cavity quantum materials can be naturally reformulated in terms of Keldysh field theory, which captures the photon statistics of emitted light as a response of driven-dissipative steady states of light and matter inside the cavity \cite{sieberer2016keldysh,flores2023non,talkington2024linear,talkington2025}. Finally, an intriguing generalization regards harnessing quantum materials for few-photon frequency multiplexing and upconversion in cavities with multiple modes, with applications in quantum sensing and communication \cite{gisin2007quantum,RevModPhys.89.035002}.

\section{Acknowledgments}

M.~C. and B.~K. acknowledge support from the Alfred P. Sloan Foundation through a Sloan Research Fellowship and a startup grant from the University of Pennsylvania. S.~T. acknowledges support from the NSF under Grant No. DGE-1845298. A.~S acknowledges support from the NSF Division of Materials Research (Award No. 1905809), from the State Secretariat for Education, Research and Innovation (SERI)-funded European Research Council Consolidator Grant TuneInt2Quantum (Award No. 101043957).

%


\appendix

\section{Raman Interactions in Cavity-Embedded Mott Insulators}
\label{app:cavityMott}

Embedded inside a high finesse optical cavity, cavity Mott insulators can be minimally described by a Hubbard model of half-filled transition-metal orbitals, which are coupled to a pair of $q=0$ TE modes $\A{\mu}$ with polarization $\mu$ via the Peierls substitution
\begin{align}
	\Ham &= \sum_{ij\sigma} t_{ij} e^{i g ( \mathbf{r}_{ij} \cdot \aVec + \aVecD \cdot \mathbf{r}_{ij} )} \CD{i\sigma} \C{j\sigma} + U \sum_i \ND{i\uparrow} \ND{i\downarrow} + \omega_0 \aVecD \aVec   \label{eq:CavityMottHamiltonian}
\end{align}
Here, $t_{ij}$ are hopping matrix elements between the valence orbitals of neighboring ions $i$ and $j$, and $\mathbf{r}_{ij}$ denote dimensionless bond vectors (in units of the lattice constant). The vectorial cavity photon operators $\aVec \equiv \sum_\mu \polVec_\mu \A{\mu}$ are defined via polarization vectors $\polVec_\mu$ and $\omega_0$ is the cavity photon frequency. The Hubbard interaction $U$ describes strong Coulomb repulsion for pairs of electrons on the same site. The dimensionless electron-photon coupling $g$ depends on the cavity geometry and scales $\sim 1/\sqrt{\mathcal{V}}$ with the mode volume. For 2D materials, coupling to TM modes can be neglected.

In the Mott phase, $U$ exceeds the electronic bandwidth and localizes a single electron per site to form a spin-1/2 magnetic moment. If the optical cavity mode with frequency $\omega_0$ is sufficiently red-detuned from the Mott gap ($\omega_0 < U$), then charge excitations remain approximately inert even for weakly pumped cavities and can be integrated out via a Schrieffer-Wolff transformation $\Ham \to \hat{P} e^{-\hat{s}} \Ham e^{\hat{s}} \hat{P}^\top$ that perturbatively decouples spin and charge dynamics ($e^{\hat{s}}$) and projects out (via a projection operator $\hat{P}$) double occupied sites \cite{sentef20}. 

To second order in a strong coupling expansion in $\sim t/U$, the coupling of spins and photons can be derived from considering individual bonds: In effect, an electron virtually hops to a neighboring site that is already occupied by an electron with opposite spin. The electron can potentially absorb a photon, however incurs a virtual intermediate-state energy penality $\sim U$. Subsequently, the second electron hops back to the initial site, realizing a photon-assisted spin exchange process. For two sites $i$ and $j$ connected via a hopping matrix element $t_{ij}$, the bond Hamiltonian can be written as
\begin{align}
    \Ham_{ij} &= -t_{ij} \sum_{\sigma} \left( e^{i \eta ( \AD{} + \A{} )} \CD{i\sigma} \C{j\sigma} + \hc \right) \notag\\
        &+ U \left( \ND{i\uparrow} \ND{i\downarrow} + \ND{j\uparrow} \ND{j\downarrow} \right) + \omega_0~ \AD{} \A{}
\end{align}
where $\eta = g |\mathbf{r}_{ij}|$ is the dimensionless light-matter coupling, $\mathbf{r}_{ij}$ is the bond vector, and $\A{}$ are \textit{rotated} cavity photon operators with bond-parallel linear polarization
\begin{align}
    \A{} = \sum_\mu \frac{\mathbf{r}_{ij} \cdot \mathbf{e}^\mu}{|\mathbf{r}_{ij}|} \A{\mu}
\end{align}
which obey $[ \A{}, \AD{} ] = 1$.

A strong coupling expansion now integrates out charge excitations, which yields a spin-exchange Hamiltonian to second order in hopping and coupled to the cavity photon modes. We start by writing Peierls couplings to photons in photon number basis:
\begin{align}
    G_{mn}&(\eta) = \bra{m} e^{i \eta ( \AD{} + \A{} )} \ket{n} \notag\\
    &= \bra{m} e^{i \eta \A{}} e^{i \eta \AD{}} e^{i \eta ( \AD{} + \A{} )} \ket{n} e^{-\frac{(i \eta)^2}{2} [\A{}, \AD{}]} \notag\\
    &= \sum_{k,l\geq 0} \frac{(i \eta)^{k+l} \bra{m} (\A{})^k (\AD{})^{l} \ket{n}}{k!l!} e^{\frac{\eta^2}{2}} \notag\\
    &= \sum_{k,l\geq 0} \frac{(i \eta)^{k+l} e^{\frac{\eta^2}{2}}}{k!l!} \sqrt{\frac{(m+k)!(n+l)!}{m!n!}} \delta_{m+k,n+l} \notag\\
    &= (i\eta)^{n-m} \sqrt{\frac{m!}{n!}} ~L_m^{n-m}(\eta^2) e^{-\eta^2/2}
\end{align}
This expansion is defined for $m \geq n$ where are $L_m^{n-m}$ the generalized Laguerre polynomials; following the same procedure for $m < n$, one finds that $G_{mn} = G_{nm}$. The hopping Hamiltonian can now be written as
\begin{align}
    \Ham = -\sum_{ij\sigma} t_{ij}  \sum_{m,n} \ketbra{m}{n} G_{mn}(\eta) ~\CD{i\sigma} \C{j\sigma}
\end{align}
where $\ket{m}$ are number eigenstates of the bond-parallel polarization-rotated cavity photon mode. To second order in hopping, a strong coupling expansion now results in a photon-dressed spin-exchange Hamiltonian
\begin{align}
    \Ham_{\textrm{eff}} &= \sum_{ij} \sum_{m,n} \mathcal{J}_{ij}^{mn} \ketbra{m}{n} \left( \hat{\mathbf{S}}_i \cdot \hat{\mathbf{S}}_j - \frac{1}{4} \right)
\end{align}
where
\begin{align}
    \mathcal{J}_{ij}^{mn} &= \sum_{l \geq 0} \left[ G_{n,l}(-\eta) G_{l,m}(\eta) + G_{n,l}(\eta) G_{l,m}(-\eta)\right] \notag\\
    &\times \left[ \frac{1}{U + (l-m) \omega_0} + \frac{1}{U + (l-n) \omega_0} \right]
\end{align}

We can now read off an expansion in terms of cavity photon operators. Writing the exchange Hamiltonian as
\begin{align}
    \Ham_{\textrm{eff}} &= \sum_{ij} \left( \hat{\mathbf{S}}_i \cdot \hat{\mathbf{S}}_j - \frac{1}{4} \right) \left[ J_{ij} + D_{ij} \AD{} \A{} + \right. \notag\\
    &\left. +~ \left( Q_{ij} \AD{} \AD{} + \hc \right) + K_{ij} \AD{} \AD{} \A{} \A{} + \dots \right]
\end{align}
one finds by inspection that
\begin{align}
    J_{ij} &= \mathcal{J}_{ij}^{0,0} \\
    D_{ij} &= \mathcal{J}_{ij}^{1,1} - J_{ij} \\
    Q_{ij} &= \frac{1}{\sqrt{2}} \mathcal{J}_{ij}^{2,0} \\
    K_{ij} &= \frac{1}{2} \left( \mathcal{J}_{ij}^{2,2} - J_{ij} - 2 D_{ij} \right) \\
    \cdots &= \cdots
\end{align}
Here, $J_{ij}$ parameterizes spin-exchange couplings dressed by vacuum fluctuations of the photon field; $D_{ij}$ describes Raman scattering of cavity photons off magnetic excitations in the material; $Q_{ij}$ describes counter-rotating spin-exchange terms that spontaneously generate or annihilate pairs of photons and are negligible provided that the cavity mode frequency is much larger than spin-exchange couplings; and $K_{ij}$ describes a higher-order non-linear \textit{two-photon} Raman scattering process.

\begin{figure}
    \centering
    \includegraphics[width=\columnwidth,trim=0cm 5cm 0cm 5cm,clip]{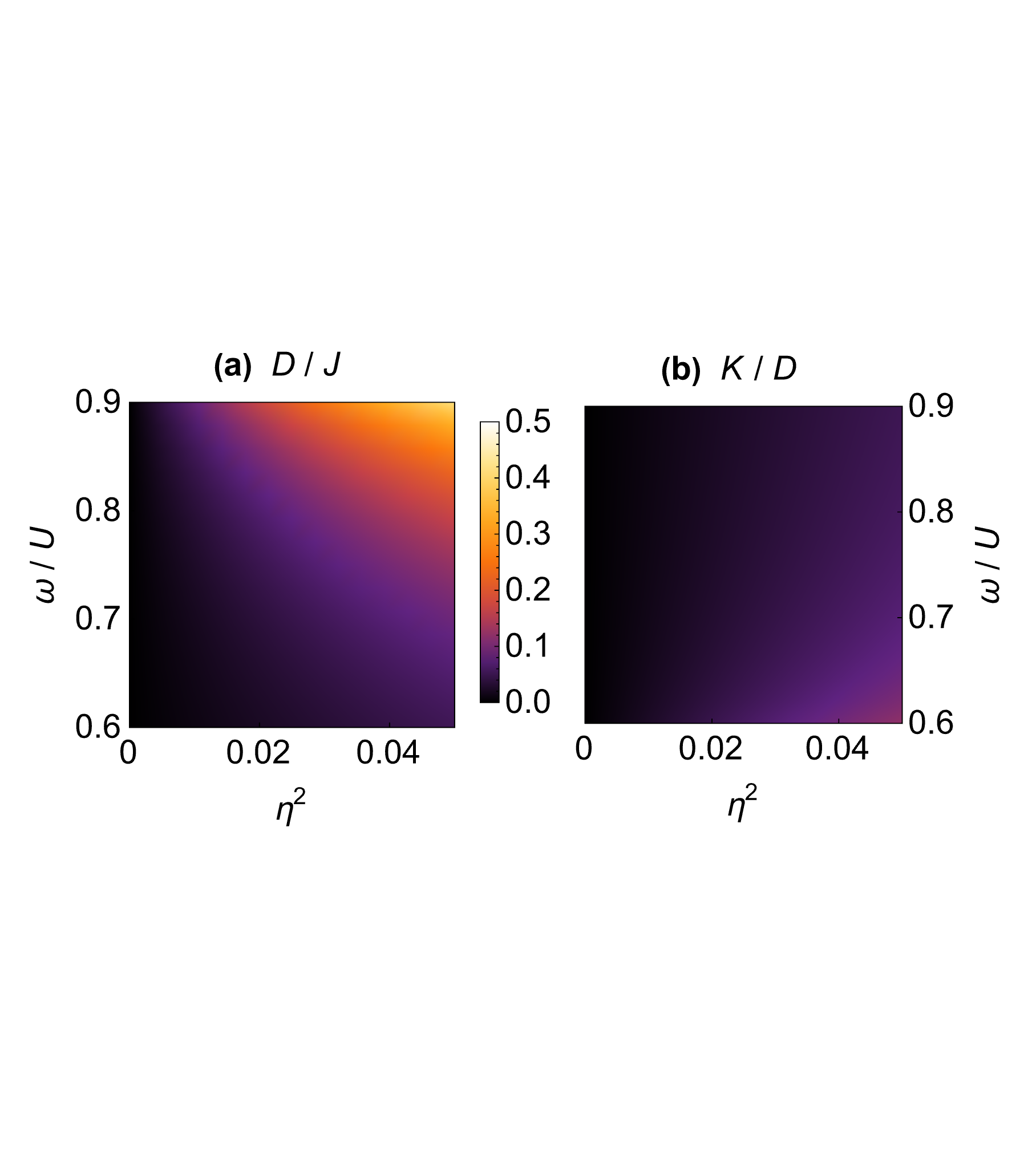}
    \caption{(a) Ratio of the spin-photon Raman exchange interaction $D \mathbf{S} \cdot \mathbf{S} ~ \AD{} \A{}$ to the nearest-neighbor Heisenberg exchange $J$, as a function of the bare cavity frequency $\omega$ in units of the Hubbard interaction $U$ and the effective light-matter coupling $\eta \equiv g r_{ij}$. (b) Ratio of the leading-order non-linear photon dressing of spin exchange $K$ to the Raman exchange interaction $D_{ij}^{\mu\nu}$. Away from deep ultrastrong light-matter coupling limits, the single-photon Raman exchange dominates over the non-linear photon dressing; the leading contribution to photon non-linearities hence arises from scattering off magnetic quantum fluctuations.}
    \label{fig:exchangeCouplings}
\end{figure}

Rotating back to the original polarization basis, and neglecting counter-rotating pair-photon generation processes as described above, one recovers an SU(2) spin-rotation invariant version of the spin-photon Hamiltonian in the main text
\begin{align}
    \Ham_{\textrm{eff}} &= \sum_{ij} J_{ij} \hat{\mathbf{S}}_i \cdot \hat{\mathbf{S}}_j + \omega_0 \sum_\mu \AD{\mu} \A{\mu} \notag\\
    &+ \sum_{ij,\mu\nu} D_{ij,\mu\nu} \left( \hat{\mathbf{S}}_i \cdot \hat{\mathbf{S}}_j - \frac{1}{4} \right) \AD{\mu} \A{\nu} ~+~ \sum_{ij} K_{ij} \cdots 
\end{align}
For instance, the polarization-dependent matrix element for the Raman scattering process reads
\begin{align}
    D_{ij}^{\mu\nu} &= \frac{\left(\mathbf{r}_{ij} \cdot \mathbf{e}^\mu\right) \left(\mathbf{r}_{ij} \cdot \mathbf{e}^\nu\right)}{|\mathbf{r}_{ij}|^2} ~ D_{ij}(\eta \to g|\mathbf{r}_{ij}|)
\end{align}
Expanding in small $g$ recovers the Fleury-Loudon formula for spin-photon couplings presented in the main text:
\begin{align}
    D_{ij}^{\mu\nu} &= 4g^2 \left(\mathbf{r}_{ij} \cdot \mathbf{e}^\mu \right) \left( \mathbf{r}_{ij} \cdot \mathbf{e}^\nu \right) t_{ij}^2 \left[ \frac{1}{U-\omega_0} + \frac{1}{U+\omega_0} - \frac{2}{U} \right]
\end{align}
However, we note that the spin-photon couplings can be resummed explicitly to all orders in $g$ via incomplete Gamma functions $\Gamma(a, b)$. For the Heisenberg exchange couplings dressed by vacuum fluctuations, one finds
\begin{align}
    J_{ij} &= \frac{4 e^{-\eta^2} (-\eta^2)^{-\frac{U}{\omega_0}} \left[ \Gamma(\tfrac{U}{\omega_0}) - \Gamma(\tfrac{U}{\omega_0},-\eta^2) \right]}{\omega_0}
\end{align}
The spin-photon Raman interactions formally become:
\begin{align}
    D_{ij} &= \frac{4}{U-\omega} \left\{ \eta^2 - 1 + \frac{U}{\omega} -e^{-\eta ^2} \left(-\eta ^2\right)^{-\frac{U}{\omega }} \right. \notag\\
        &\times \left[\eta ^2 \left(\eta
    ^2-2\right)+\frac{U^2}{\omega ^2}+\frac{\left(2 \eta ^2-1\right) U}{\omega
    }\right] \notag\\
    &\times \left. \left[\Gamma \left(\frac{U}{\omega }\right)-\Gamma \left(\frac{U}{\omega
    },-\eta ^2\right)\right] \right\}
\end{align}
while higher-order two-photon Raman couplings (quartic on photon operators) become:
\begin{align}
    K_{ij} &= \frac{1}{(U-\omega)(U-2\omega)} \left\{ \frac{U^4}{\omega^3} + e^{-\eta ^2} \left(-\eta ^2\right)^{-\frac{U}{\omega }} \times \right. \notag\\
    &\times \left[\eta ^4 \
    \left(\eta ^4-8 \eta ^2+12\right) \omega +2 \left(2 \eta ^2-1\right) \frac{U^3}{\omega^2} \right. \notag\\
    &+\left. \left(6 \eta ^2 \left(\eta ^2-2\right)-1\right) \frac{U^2}{\omega} + 2 \left(2 \eta ^6-9 \eta ^4+4 \eta ^2+1\right) U\right] \notag\\
    &\times \left[\Gamma \left(\frac{U}{\omega }\right)-\Gamma \
    \left(\frac{U}{\omega },-\eta ^2\right)\right] - \omega \notag\\
    &\times \left[\left(\eta ^6-7 \eta ^4+6 \eta ^2+2\right) + \frac{U^3}{\omega^3} \right.\notag\\
    &+ \left. \left. \left(3 \eta ^2-2\right) \frac{U^2}{\omega^2} +\left(3 \eta ^4-9 \eta \
    ^2-1\right) \frac{U}{\omega} \right] \right\}
\end{align}
To ascertain that the two-photon Raman scattering processes ($K_{ij}$) are negligible for optical cavities, the ratios $D/J$ and $K/D$ are shown in Fig. \ref{fig:exchangeCouplings}, as a function of the detuning of the bare cavity frequency from the Mott gap $\omega_0/U$ and the squared light-matter coupling strength $\eta^2$. While single-photon Raman processes can lead to a pronounced renormalization of the spin-exchange couplings for strong light-matter interactions, importantly, two-photon Raman processes are strongly suppressed for the optical cavities provided that two-photon resonances with the charge gap are avoided. Consequently, the leading contribution to photon non-linearities in the cavity Mott insulator arises from scattering off magnetic quantum fluctuations in the material.

We note that, in principle, the cavity photon operators transform as well, acquiring a small matter component upon integrating out charge excitations. For instance, after the Schrieffer-Wolff transformation and projecting out charge excitations, the cavity photon number operator (ignoring polarization) transforms as
\begin{align}
    \hat{P} e^{-\hat{s}} ~\AD{}\A{}~ e^{\hat{s}} \hat{P}^\top  \approx \AD{} \A{} + \sum_{ij} M_{ij} \hat{\mathbf{S}}_i \cdot \hat{\mathbf{S}}_j + \cdots
\end{align}
where
\begin{align}
    M_{ij} = 4t_{ij}^2 (r_{ij} g)^2 \left[ \frac{1}{(U+\omega_0)^2} - \frac{1}{(U-\omega_0)^2} \right] + \mathcal{O}(g^4)
\end{align}
However, their dark-cavity vacuum expectation values cannot be observed outside the cavity \cite{ciuti06} in the absence of a drive. With driving, upon injecting real photons into the cavity, these contributions enter as higher-order corrections to the Raman scattering processes ($D_{ij}$) and can thus be neglected.

\section{Input-Output Relations for Cavity Quantum Materials}
\label{app:inputOutputMott}

We now derive the input-output relation for output photons, as a function of input fields and an infinite series of scattering processes between cavity photons and elementary magnetic excitations in the material. Starting from the cavity $\Ham_{\textrm{cav}}$, bath $\Ham_{\textrm{B}}$, and interaction $\Ham_{\textrm{int}}$ Hamiltonians, with $\Ham_{\textrm{cav}}$ as in Eq. (\ref{eq:CavitySpinPhotonExchangeModel}) of the main text and

\begin{align}
    \Ham_{\textrm{B}} &= \int d\omega ~ \omega ~ \bVecXD{\alpha}(\omega; t) \cdot \bVecX{\alpha}(\omega; t)\\
    \Ham_{\textrm{int}} &= \sqrt{\frac{\gamma}{2\pi}}\int d\omega ~  \bVecXD{\alpha}(\omega; t) \cdot \aVec(t)+\aVecD(t) \cdot \bVecX{\alpha}(\omega; t)
\end{align}
the Heisenberg equations of motion for cavity photons reads
\begin{align}
    &\partial_t \aVec(t) = i\left[ \Ham_{\textrm{cav}} + \Ham_{\textrm{B}} + \Ham_{\textrm{int}},~ \aVec(t) \right] \notag\\
        &= -i\left[ \omega_0 - i\gamma + \Dop(t) \right] \cdot \aVec(t) - i\sqrt{\frac{\gamma}{2\pi}} \sum_{\alpha} \int d\omega~ \bVecX{\alpha}(\omega; t)
\end{align}
where $\alpha = L, R$ indexes left- and right-side free space photon modes, and $\bVecX{\alpha}(\omega; t)$ are free space photon fields at Heisenberg time $t$. Their equations of motion read
\begin{align}
    \partial_t \bVecX{\alpha}(\omega; t) &= -i\omega~ \bVecX{\alpha}(\omega; t) - i\sqrt{\frac{\gamma}{2\pi}} \aVec(t)
\end{align}
Denoting $t_i \to -\infty$ the initial time of the experiment, this equation can be integrated
\begin{align}
    \bVecX{\alpha}(\omega; t) &= e^{-i\omega (t-t_i)} \bVecX{\alpha}(\omega; t_i) - i\sqrt{\frac{\gamma}{2\pi}} \int\limits_{t_i}^{t} d\tau e^{-i\omega (t-\tau)} \aVec(\tau)
\end{align}
Substituting into the Heisenberg equation of motion for the cavity photons, one finds
\begin{align}
    \partial_t \aVec(t) &= -i\left[ \omega_0 - i\gamma + \Dop(t) \right] \cdot \aVec(t) \notag\\
        &- i \sqrt{\gamma} \sum_\alpha \int \frac{d\omega}{\sqrt{2\pi}}~ e^{-i\omega (t-t_i)} \bVecX{\alpha}(\omega; t_i) \notag\\
        &- \frac{\gamma}{\pi} \int\limits_{t_i}^{t} d\tau \int d\omega e^{-i\omega(t-\tau)} \aVec(\tau)
\end{align}
Taking $t_i \to -\infty$, the third line evaluates to a Dirac $\delta$ function by virtue of the frequency-independent photon tunnelling $\sqrt{\gamma/2\pi}$. Using the definition of the input field $\bInVecX{L}(t) \equiv (1/\sqrt{2\pi}) \int d\omega e^{-i\omega(t-t_i)} \bInVecX{L}(\omega; t_i)$ one arrives at the cavity photon equation of motion [Eq. (\ref{eq:cavityEoM})] presented in the main text.

\section{Input-Output Relation for Second-Order Photon Coherence}
\label{app:inputOutputG2}

\begin{figure*}
    \centering
    \includegraphics[width=\textwidth]{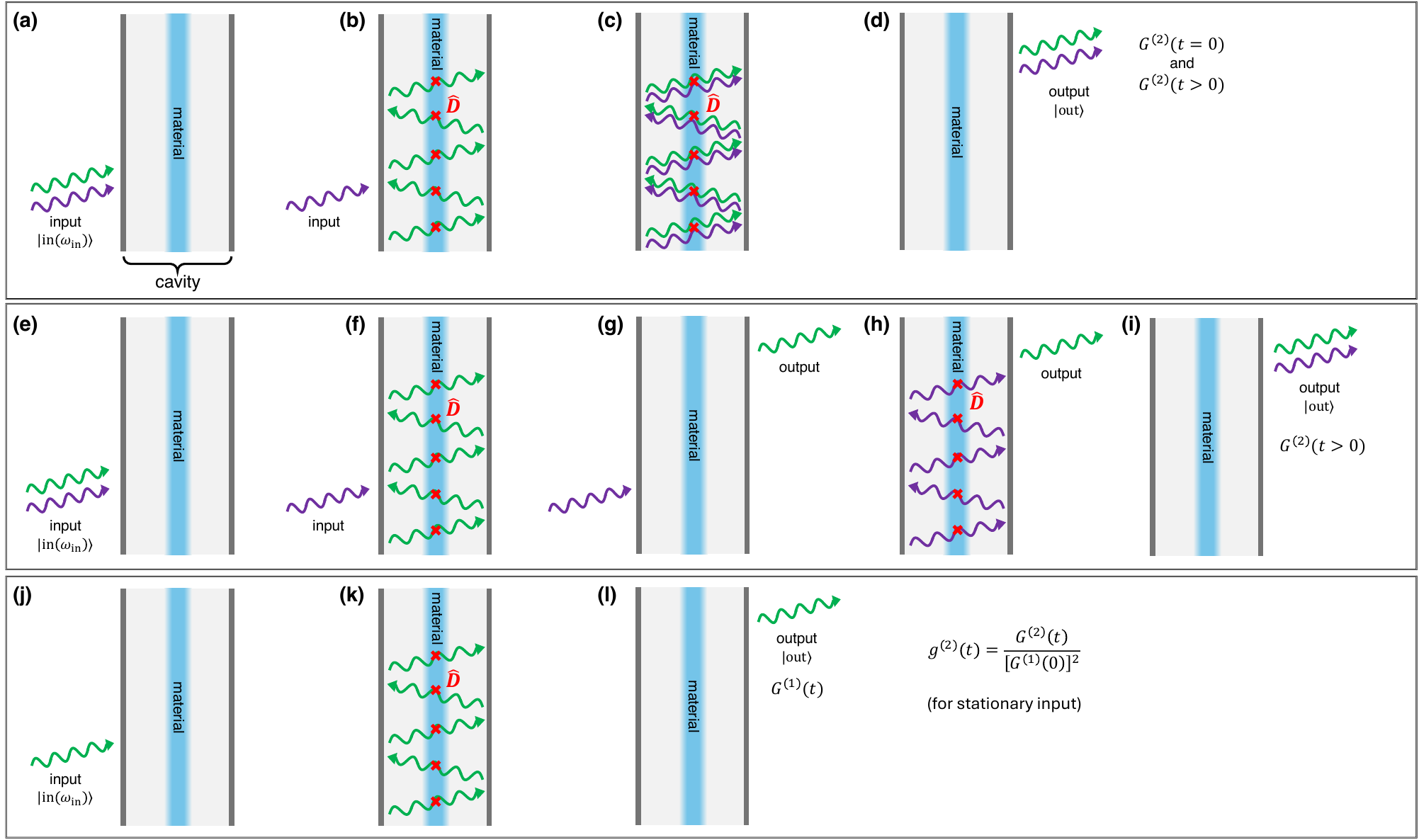}
    \caption{\textbf{Schematics of photon tunneling processes for $g^{(2)}(t)$ in the strong light-matter coupling regime.} (a)-(d) Pair-photon tunneling process: Starting from (a) an arbitrary left-side input state, one photon tunnels (b) into a high quality cavity and dresses the material via repeated light-matter scattering processes $\Dop$. Then, (c) a second photon tunnels in; the material therefore gets dressed by two photons, while potentially rotating the photon polarizations. Finally, both photons (d) tunnel out to the right and are detected. This process contributes to the two-photon correlation function $G^{(2)}(t) = \expect{ \bOutVecD(0) \bOutVecD(t) \bOutVecD(t) \bOutVec(0) }$ both at coincidence ($t=0$) and finite times ($t>0$. (e)-(i) Sequential two-photon tunneling process: Again starting from (e) a left-side input state, one photon tunnels (f) into the cavity, dresses the material, but (g) leaves the cavity towards the detector before a second photon tunnels in. (h) Subsequently, a second photon tunnels into the cavity, and (i) tunnels out towards the detector and is detected as well. This processes contributes to the two-photon correlation function at finite time delays $t>0$. (j)-(l) Single-photon tunneling process: Starting from (j) an arbitrary input state, one photon tunnels (k) into the cavity, dresses the material, and (l) leaves the cavity and is detected. Single-photon tunneling determines the single-photon correlation function $G^{(1)}(t) = \expect{ \bOutVecD(t) \bOutVec(0)}$. For high quality cavities and weak input fields, two-photon corrections to $G^{(1)}(t)$ can be neglected for purposes of computing the second-order photon coherence $g^{(2)}(t) = G^{(2)}(t) / [ G^{(1)}(0) ]^2 $ for stationary (monochromatic) input fields.}
    \label{fig:g2processes}
\end{figure*}

In this section, we derive the input-output relation for the second-order photon coherence $g^{(2)}(t)$ as a function of the input field and an infinite series of scattering processes of cavity photons and elementary magnetic excitations in the material. For stationary monochromatic input fields, the second-order coherence can be written as a ratio
\begin{align}
    g^{(2)}(t) = G^{(2)}(t) / [ G^{(1)}(0) ]^2  \label{eq:g2fromG1G2}
\end{align}
of two-photon [$G^{(2)}(t)$] and single-photon [$G^{(1)}(t)$] correlators for light transmitted through the cavity.

\subsection{Single-photon output correlation function}

Explicitly including the dependence on the polarization of input photons as well as the polarization of transmitted photons at the detector and using Eq. (\ref{eq:inputOutput}) in the main text, the single-photon correlation function reads
\begin{align}
    &G^{(1)}_{\mu}(t-t') = \expectNormalOrder{ \BoutXD{R,\mu}(t) \BoutX{R,\mu}(t') } \\
        &\hspace{1.7cm}= \gamma^2 \int\limits_{0}^{\infty} d\tau d\tau' e^{i\omega_0(\tau-\tau') - \gamma(\tau+\tau')} ~\times\notag\\
        &\times  \expect{ \left[ \bVecXD{L}(t-\tau) \Uop^\dag(t,t-\tau) \polVec_\mu \right] \left[ \polVec^*_\mu \Uop(t',t'-\tau') \bVecX{L}(t'-\tau') \right] }
\end{align}
where $\Uop(t,t')$ is the time-ordered propagator defined in Eq. (\ref{eq:scatteringOperator}), and subscripts $\textrm{in}$ are dropped for input photons, for convenience. As described in the main text, the photon-dressed spin-exchange operator $\Dop$ that enters in the time-ordered propagator can now be approximated to evolve solely under the Hamiltonian $\Ham$ of the material
\begin{align}
    \Uop(t,t') &= \TimeOrder \exp\left\{ -i \int_{t'}^{t} d\tau ~\Dop(\tau) \right\} \notag\\
    &\approx e^{i \Ham t} e^{i ( \Ham + \Dop ) (t-t')} e^{-i\Ham t'}  \label{eq:scatteringOperatorApproximation}
\end{align}
Note that $\Dop$ and $\Uop$ are $2 \times 2$ matrices of operators in the photon polarization basis; the exponential of $\Ham + \Dop$ must therefore be understood accordingly. Similarly, $\polVec_\mu \Uop \bVecX{L}$ denotes an operator-valued vector-matrix-vector product in polarization basis. $G^{(1)}$ now factorizes into an expectation value for input fields times a correlation function for the material. Furthermore, for a monochromatic input field with frequency $\wIn$, the input fields simplify to
\begin{align}
    \bInVecX{L}(t) \to \bInVecX{L}(\wIn) e^{-i\wIn t}  \label{eq:monochromaticInput}
\end{align}
In this case, $G^{(1)}$ is stationary and the dependence on $t+t'$ vanishes; shifting the time argument $t-t' \to t$, the one-photon correlation function can now be written succinctly as
\begin{align}
    G^{(1)}_{\mu}(t) &= \gamma^2 \sum_{\nu\nu'} \int\limits_{0}^{\infty} d\tau d\tau' e^{-i\wIO(\tau-\tau') - \gamma(\tau+\tau')} n_{\nu\nu'} e^{i\wIn t} \notag\\
    &\times \expect{ \left[ \polVec^*_\mu \Uop(t,t-\tau) \polVec_\nu \right]^\dag \left[ \polVec^*_\mu \Uop(0,-\tau') \polVec_\nu \right] } 
\end{align}
where $\wIO \equiv \wIn - \omega_0$, and $n_{\nu\nu'}$ is the one-photon expectation value for the input field that depends on the polarization of the input field
\begin{align}
    n_{\nu\nu'} = \expect{ \BinXD{L\nu}(\wIn) \BinX{L\nu'}(\wIn) } \equiv \sum_j \bar{f}_{\nu}^{(j)} f_{\nu'}^{(j)}
\end{align}
which can be decomposed into a set of form factors $f_\nu^{(j)}$ with $j$ indexing the irreducible representations for the photon field. Here, $\bar{f}$ denotes complex conjugation. For instance, for materials with four-fold rotation symmetry, it is useful to work with $j \in [ A_{1\textrm{g}},A_{2\textrm{g}},B_{1\textrm{g}},B_{2\textrm{g}} ]$. Now, inserting a complete set of final states $\ket{n}$ for the material, the time integrals can be performed explicitly, and one arrives at
\begin{align}
    G_\mu^{(1)}(t) = \gamma^2 \sum_{n,j} &\left| \bra{n} \polVec^*_\mu \frac{1}{\wIO + E_0 - \Ham - \Dop + i\gamma} \fVec{j} \ket{0} \right|^2 \notag\\
        &\times e^{i(\wIn + E_0 - E_n) t}    \label{eq:G1complete}
\end{align}
where $\fVec{j} \equiv \sum_\nu \polVec_\nu f_\nu^{(j)}$ and $\ket{0}$ is the material ground state. Note that the Fourier transform of $G^{(1)}(t)$ takes the form of a Fermi golden rule for transmitting a single photon through the cavity. Importantly, the resolvent operator in the above equation again must be understood as the inverse of a matrix of operators. It can be evaluated using standard block matrix inversion identities. Conversely, for quasi-one-dimensional systems where e.g. only photons polarized in the $x$ direction couple to the material, the above expression readily simplifies via $\Dop \to \hat{D} \equiv \hat{D}_{xx}$.

\subsection{Two-photon output correlation function}

The two-photon output correlation function $G^{(2)}(t)$ can be evaluated in a similar manner. Inserting a complete set of states $\ket{n}$ into $G^{(2)}(t)$, one finds
\begin{align}
    G_{\mu\mu'}^{(2)} &= \expectNormalOrder{ \BoutXD{R,\mu'}(t') \BoutXD{R,\mu}(t) \BoutX{R,\mu}(t) \BoutX{R,\mu'}(t') } \notag\\
    &\equiv \gamma^4 \sum_n \braket{\Psi_{\mu\mu'}(t,t')}{n} \braket{n}{\Psi_{\mu\mu'}(t,t')}
\end{align}
where $\ket{\Psi_{\mu\mu'}(t,t')}$ is defined as
\begin{align}
    \ket{\Psi_{\mu\mu'}(t,t')} &= \TimeOrder \A{\mu}(t) \A{\mu'}(t') \ket{i} \\
    &= \int\limits_{-\infty}^t d\tau \int\limits_{-\infty}^{t'} d\tau' e^{i(\omega_0-i\gamma)(\tau+\tau'-t-t')} ~\times \notag\\
    &\hspace{-1cm}\times \TimeOrder \left[ \polVec^*_\mu \Uop(t,\tau) \bVecX{L}(\tau) \right] \left[ \polVec^*_{\mu'} \Uop(t',\tau') \bVecX{L}(\tau') \right] \ket{i}
\end{align}
Here, $\ket{i} \equiv \ket{\rm in} \otimes \ket{0}$ is an initial product state of an arbitrary input field $\ket{\rm in}$ and the material ground state $\ket{0}$. We take $t \geq t'$ without loss of generality; $G_{\mu\mu'}^{(2)}(t,t')$ is symmetric under exchange $t \leftrightarrow t'$ and $\mu \leftrightarrow \mu'$.

In contrast to $G^{(1)}$, special care must be taken to properly interpret the time-ordered product of the two operator-valued $2\times 2$ matrices $\Uop(t,\tau)$ and $\Uop(t',\tau')$. Whereas the space of polarizations for a single photon in $G^{(1)}$ is two-dimensional, the expression for $G^{(2)}$ can be understood as summing up two processes: (I) the sequential tunneling of a single photon through the cavity, followed by a second photon; and (II) the tunneling of \textit{two} photons into the cavity (yielding a two-photon intermediate state), followed by their sequential tunneling out. Both processes are depicted schematically in Fig. \ref{fig:g2processes}. In process (I), the intermediate-state dynamics within the cavity span a product space of the material Hilbert space and the two-dimensional space of the polarization of a single cavity photon, in analogy to $G^{(1)}$. However, for process (II), the intermediate cavity state with two photons must span a product space of the material Hilbert space and the \textit{four}-dimensional tensor product space of the polarizations of two cavity photons.

To illustrate this explicitly, we evaluate the final state $\ket{\Psi_{\mu\mu'}(t,t')}$ using the input-output equations of motion while explicitly keeping track of time ordering. First, assuming $t \geq t'$ the cavity photon $\A{\mu}(t)$ can be evolved to time $t'$ via the equation of motion [Eq. (\ref{eq:cavityEoM})] which integrates to
\begin{align}
    \A{\mu}(t) &= e^{-i\Omega (t-t')} \polVec_\mu^* \Uop(t,t') \aVec(t') \notag\\
        &- i\sqrt{\gamma} \int\limits_{t'}^{t} d\tau~ e^{-i\Omega(t-\tau)} \polVec_\mu^* \Uop(t,\tau) \bVecX{L}(\tau)
\end{align}
with $\Omega \equiv \omega_0 - i\gamma$, from which one obtains
\begin{align}
    \ket{\Psi_{\mu\mu'}} &= e^{-i\Omega(t-t')} \sum_{\rho} \polVec_\mu^* \Uop(t,t') \polVec_\rho \hat{P}_{\rho\mu'}(t') \notag\\
        &- i\sqrt{\gamma} \int\limits_{t'}^{t} d\tau~ e^{-i\Omega(t-\tau)} \polVec_\mu^* \Uop(t,\tau) \bVecX{L}(\tau) \A{\mu'}(t') \ket{0}
\end{align}
where
\begin{align}
    \hat{P}_{\rho\mu'}(t') \equiv \A{\rho}(t') \A{\mu'}(t')
\end{align}
is a photon pair annihilation operator at time $t'$. The input field $\bVecX{L}(\tau)$ can be moved to the right of $\A{\mu'}(t')$ by virtue of the causality relations since $\tau \geq t'$. The second line describes the sequential single-photon tunneling process process (I). $\A{\mu'}(t')$ can be expressed in terms of the input field by evolving from time $t_i \to -\infty$ via
\begin{align}
    \A{\mu'}(t') = -i\sqrt{\gamma} \int\limits_{-\infty}^{t'} d\tau'~ e^{-i\Omega(t'-\tau')} \polVec_{\mu'}^* \Uop(t',\tau') \bVecX{L}(\tau')
\end{align}

Conversely, relating the two-photon annihilation operator $\hat{P}_{\rho\mu'}(t')$ to the input fields needs more care, as its equation of motion operates in a product space of polarizations for both photons. Since bosons commute, the two-photon annihilation operator is symmetric under permutation of the polarization indices, obeying $\hat{P}_{\rho\mu'}(t') = \hat{P}_{\mu'\rho}(t')$. Its equation of motion reads
\begin{align}
    \partial_t P_{\mu\mu'}(t) &= -2i(\omega_0 - i\gamma) P_{\mu\mu'}(t) \notag\\
        &- i \sum_{\nu} \left( \hat{D}_{\mu\nu}(t) P_{\nu\mu'}(t) + \hat{D}_{\mu'\nu}(t) P_{\nu\mu}(t) \right)
\end{align}
where $\hat{D}_{\mu\nu} \equiv \polVecC_\mu \cdot \Dop \cdot \polVec_\nu$. In principle, there are \textit{three} distinct polarization states. For instance, working in a basis of linear polarizations $x$ and $y$, these read $\ket{x,x}$, $\ket{y,y}$ and $\ket{x,y} (= \ket{y,x})$. However, it will instead be formalistically advantageous to instead work in an overcomplete \textit{four-state} polarization basis (e.g., $\ket{x,x}$, $\ket{y,y}$, $\ket{x,y}$, $\ket{y,x}$), and implicitly ensure $\ket{x,y} = \ket{y,x}$ as a symmetry of the resulting input-output relations. This approach is exact. We denote a four-dimensional \textit{vector} of two-photon annihilation operators $\hat{\mathbf{P}}(t)$ in this four-dimensional space of two-photon polarizations, and define its operator-valued elements as $[\hat{\mathbf{P}}]_{(\rho\mu')}(t') \equiv \hat{P}_{\rho\mu'}$ where $(\rho\mu')$ indexes the two-photon polarization state. For instance, for a basis of linear polarizations one would obtain
\begin{align}
    \hat{\mathbf{P}}(t) = \left[\begin{array}{c} \hat{P}_{xx}(t) \\ \hat{P}_{xy}(t) \\ \hat{P}_{yx}(t) \\ \hat{P}_{yy}(t) \end{array}\right]
\end{align}
The Heisenberg equation of motion for $\hat{\mathbf{P}}$ takes a succinct form
\begin{align}
    \partial_{t} \hat{\mathbf{P}} &= -i \left[ 2\Omega + \Dop(t) \oplus  \Dop(t) \right] \hat{\mathbf{P}}(t) \notag\\
        &- i \sqrt{\gamma} \left[ \aVec(t) \otimes \bVecX{L}(t) + \bVecX{L}(t) \otimes \aVec(t) \right]
\end{align}
Here, $\Dop \oplus \Dop$ denotes the Kronecker sum $\Dop \otimes \mathbf{1} + \mathbf{1} \otimes \Dop$, and $\mathbf{1}$ is the dimension-two identity matrix for one-photon polarizations. The equation can be again formally integrated. After permuting the tensor products, one finds:
\begin{align}
    &\hat{P}_{\rho\mu'}(t') = -i\sqrt{\gamma} \int_{-\infty}^{t'} d\tau e^{-2i\Omega(t'-\tau)} ~\times \notag\\
    &~~~\times \polTensorC_{\rho\mu'} \cdot \left[ \Uop(t',\tau) \oplus \Uop(t',\tau) \right] \cdot \left[ \aVec(\tau) \otimes \bVecX{L}(\tau) \right]
\end{align}
where
\begin{align}
    \polTensor_{\rho\mu'} = \polVec_\rho \otimes \polVec_{\mu'} + \polVec_{\mu'} \otimes \polVec_{\rho}
\end{align}
is the symmetrized two-photon polarization vector. Substituting into the expression for $\ket{\Psi_{\mu\mu'}}$, one finally arrives at
\begin{align}
    \ket{\Psi_{\mu\mu'}} &= -\gamma^2 \left( \ket{\Psi_{\mu\mu'}^{(1)}} + \ket{\Psi_{\mu\mu'}^{(2)}} \right)
\end{align}
where
\begin{align}
    \ket{\Psi_{\mu\mu'}^{(1)}} &= -\gamma \sum_{\nu\nu'} \int\limits_{t'}^{t} d\tau \int\limits_{-\infty}^{t'} d\tau' e^{-i\Omega(t+t'-\tau-\tau')} \notag\\
    &\times \hat{U}_{\mu\nu}(t,\tau) \hat{U}_{\mu'\nu'}(t',\tau') \B{L\nu'}(\tau') \B{L\nu}(\tau) \ket{0}
\end{align}
describes the sequential-tunneling process (I), whereas
\begin{align}
    \ket{\Psi_{\mu\mu'}^{(2)}} &= -\gamma \sum_{\rho\rho'\nu\nu'}\int\limits_{-\infty}^{t'} d\tau \int\limits_{-\infty}^{\tau} d\tau' e^{-i\Omega(t+t'-\tau-\tau')} \notag\\
    \times& ~\hat{U}_{\mu\rho}(t,t') \left[ \hat{U}_{\rho\rho'}(t',\tau) \hat{U}_{\mu'\nu}(t',\tau) ~+ \right.\notag\\
    +& \left. \hat{U}_{\rho\nu}(t',\tau) \hat{U}_{\mu'\rho'}(t',\tau) \right] \hat{U}_{\rho'\nu'}(\tau,\tau') \B{L\nu'}(\tau') \B{L\nu}(\tau) \ket{0}
\end{align}
The time integrals can again be performed explicitly by working in a product space of the material Hilbert space and the space of photon polarizations.

Using the expression for the scattering operator [Eq. (\ref{eq:scatteringOperatorApproximation})] and considering again the case of a monochromatic stationary input field [Eq. (\ref{eq:monochromaticInput})], the response can again be written as a product of a material correlation function and an expectation value for the input field.

For process (I), we have $\hat{U}_{\mu\nu}(t,\tau) \hat{U}_{\mu'\nu'}(t',t'-\tau') \approx e^{i\Ham t} (\polVecC_{\mu} e^{-i(\Ham + \Dop) (t-\tau)} \polVec_{\nu}) e^{-i\Ham (\tau-t')} (\polVecC_{\mu'} e^{-i(\Ham + \Dop)\tau'} \polVec_{\nu'}) \times e^{-i\Ham (t'-\tau')}$. We insert a complete set of intermediate states $\ket{m}$ and find
\begin{align}
    &\braket{n}{\Psi_{\mu\mu'}^{(1)}} = -\gamma \sum_m \sum_{\nu\nu'} \bra{n} \int\limits_{t'}^{t} d\tau \int\limits_{0}^{\infty} d\tau' e^{-i\Omega(t-\tau + \tau')} \notag\\
    &\times e^{i E_n t} \left[\polVecC_\mu e^{-i(\Ham+\Dop)(t-\tau)} \polVec_\nu \right] e^{-i\Ham (\tau-t')} \ketbra{m}{m} \notag\\
    &\times \left[\polVecC_{\mu'} e^{-i(\Ham + \Dop)\tau'} \polVec_{\nu'}\right] \B{L\nu'} \B{L\nu} \ket{0} e^{-i\wIn(\tau+t'-\tau')} e^{-i E_0 (t'-\tau')}
\end{align}
Performing the integral over $\tau'$ is straightforward; with $\Omega \equiv \omega_0 - i\gamma$, one finds
\begin{align}
    &\braket{n}{\Psi_{\mu\mu'}^{(1)}} = -i\gamma \sum_m \sum_{\nu\nu'} \bra{n} \int\limits_{t'}^{t} d\tau~ e^{-i\Omega(t-\tau)} \notag\\
    &\times e^{i E_n t} \left[\polVecC_\mu e^{-i(\Ham+\Dop)(t-\tau)} \polVec_\nu \right] e^{-i E_m (\tau-t')} \ketbra{m}{m} \notag\\
    &\times \left[\polVecC_{\mu'} \frac{1}{\wIO + E_0 - \Ham + \Dop + i\gamma} \polVec_{\nu'}\right] \B{L\nu'} \B{L\nu} \ket{0} e^{-i\wIn \tau} e^{-i E_0 t'}
\end{align}
The integral over $\tau$ can now be performed and one obtains
\begin{align}
    &\braket{n}{\Psi_{\mu\mu'}^{(1)}} = \gamma \sum_m \sum_{\nu\nu'} \bra{n} e^{i (E_n - E_m) (t-t')} \notag\\
    &\times \left[\polVecC_\mu \frac{1 - e^{i(\wIO+E_m-\Ham-\Dop+i\gamma)(t-t')}}{\wIO + E_m - \Ham - \Dop + i\gamma} \polVec_\nu \right] \ketbra{m}{m} \notag\\
    &\times \left[\polVecC_{\mu'} \frac{1}{\wIO + E_0 - \Ham + \Dop + i\gamma} \polVec_{\nu'}\right] \B{L\nu'} \B{L\nu} \ket{0}
\end{align}
after discarding an overall time-dependent phase.

We now turn to process (II). Inserting the resolution of the time-ordered scattering operator as well as a set of intermediate states $\ket{m}$, one finds
\begin{align}
    &\braket{n}{\Psi_{\mu\mu'}^{(2)}} = -\gamma \sum_m \sum_{\rho\rho'\nu\nu'} \bra{n} \int\limits_{-\infty}^{t'} d\tau \int\limits_{-\infty}^{\tau} d\tau' e^{-i\Omega(t+t'-\tau-\tau')} \notag\\
    &\times ~e^{iE_n t} \left[ \polVecC_\mu e^{-i(\Ham+\Dop)(t-t')} \polVec_\rho \right] e^{-iE_m t'} \ketbra{m}{m} e^{iE_m t'} \notag\\
    &\times \left[ \polTensorC_{\rho\mu'} ~e^{-i(\Ham + \Dop \oplus \Dop)(t'-\tau)} ~(\polVec_{\rho'} \otimes \polVec_{\nu}) \right] \notag\\
    &\times \left[ \polVecC_{\rho'} e^{-i(\Ham+\Dop)(\tau-\tau')} \polVec_{\nu'} \right] e^{-iE_0 \tau'} e^{-i\wIn(\tau+\tau')} \B{L\nu'} \B{L\nu} \ket{0}
\end{align}
The time integrals can be evaluated straightforwardly and one finds
\begin{align}
    &\braket{n}{\Psi_{\mu\mu'}^{(2)}} = \gamma \sum_{\rho\rho'\nu\nu'} \bra{n} \left[ \polVecC_{\mu} e^{i(\wIO + E_n - \Ham - \Dop + i\gamma)(t-t')} \polVec_{\rho} \right] \notag\\
    &\times \left[ \polTensorC_{\rho\mu'} \frac{1}{2\wIO + E_0 - \Ham - \Dop \oplus \Dop + 2i\gamma} ~(\polVec_{\rho'} \otimes \polVec_{\nu})\right] \notag\\
    &\times \left[ \polVecC_{\rho'} \frac{1}{\wIO + E_0 - \Ham - \Dop + i\gamma} \polVec_{\nu'} \right] ~ \B{L\nu'} \B{L\nu} \ket{0}
\end{align}
after discarding the same overall time-dependent phase.

\begin{widetext}

We now decompose the two-photon input correlator
\begin{align}
    \expect{ \BinXD{L\nu_1} \BinXD{L\nu'_1} \BinX{L\nu'_2} \BinX{L\nu_2} } \equiv \sum_j \left\{ \sum_i \left[ \FVecIJ{i}{j} \right]_{\nu_1} \left[ \FVecIJ{i}{j} \right]_{\nu'_1} \right\}^* \left\{ \sum_i \left[ \FVecIJ{i}{j} \right]_{\nu'_2} \left[ \FVecIJ{i}{j} \right]_{\nu_2} \right\}
\end{align}
into a set of trajectories $j$ of semi-classical states (with $i$ an internal index) described by amplitude-weighted polarization vectors $\FVecIJ{i}{j}$. Combining processes (I) and (II) to compute $G^{(2)}(t)$, we finally arrive at the two-photon output correlation function
\begin{align}
    &G_{\mu\mu'}^{(2)}(t) = \gamma^4 \sum_{j,n} \left| \sum_i \bra{n} \left\{ \sum_m e^{i (E_n - E_m) t} \left[\polVecC_\mu \frac{1 - e^{i(\wIO+E_m-\Ham-\Dop+i\gamma) t}}{\wIO + E_m - \Ham - \Dop + i\gamma} \FVecIJ{i}{j} \right] \ketbra{m}{m} \left[\polVecC_{\mu'} \frac{1}{\wIO + E_0 - \Ham - \Dop + i\gamma} \FVecIJ{i}{j} \right] \right.\right. \notag\\
    &+ \left.\left. \sum_{\rho} \left[ \polVecC_{\mu} e^{i(\wIO + E_n - \Ham - \Dop + i\gamma)t} \polVec_{\rho} \right]
    \polTensorC_{\rho\mu'}~ \frac{1}{2\wIO + E_0 - \Ham - \Dop \oplus \Dop + 2i\gamma} ~\FVecIJ{i}{j} \otimes \left[ \frac{1}{\wIO + E_0 - \Ham - \Dop + i\gamma} \FVecIJ{i}{j} \right] \right\} \ket{0} \right|^2  \label{eq:G2complete}
\end{align}
Importantly, at coincidence $t=0$, the sequential photon tunneling process (II) vanishes; the coincidence expression for photon (anti)bunching therefore takes an appealingly simple form
\begin{align}
    &G_{\mu\mu'}^{(2)}(0) = \gamma^4 \sum_{j,n} \left| \sum_i \bra{n} \left[ \polTensorC_{\mu\mu'} \frac{1}{2\wIO + E_0 - \Ham - \Dop \oplus \Dop + 2i\gamma} \FVecIJ{i}{j} \otimes \left[ \frac{1}{\wIO + E_0 - \Ham - \Dop + i\gamma} \FVecIJ{i}{j} \right] \right] \ket{0} \right|^2
\end{align}
where $\polTensor_{\mu\mu'} = (\polVec_\mu \otimes \polVec_{\mu'} + \polVec_{\mu'} \otimes \polVec_\mu)$ is the symmetrized two-photon polarization vector. Substituting into Eq. (\ref{eq:g2fromG1G2}), we therefore arrive at a succinct expression for the polarization-resolved second-order photon coherence at coincidence:
\begin{align}
    g_{\mu\mu'}^{(2)}(0) = \frac{ \sum_{j,n} \left| \sum_i \bra{n} \left[ \polTensorC_{\mu\mu'} \frac{1}{2\wIO + E_0 - \Ham - \Dop \oplus \Dop + 2i\gamma} \FVecIJ{i}{j} \otimes \left[ \frac{1}{\wIO + E_0 - \Ham - \Dop + i\gamma} \FVecIJ{i}{j} \right] \right] \ket{0} \right|^2  }{  \prod\limits_{\sigma=\{\mu,\mu'\} } \left[ \sum\limits_{j,n} \left| \bra{n} \polVecC_\sigma \frac{1}{\wIO + E_0 - \Ham - \Dop + i\gamma} \fVec{j} \ket{0} \right|^2 \right]  }  \label{eq:g2withpolarizationAppendix}
\end{align}
The polarization-resolved second-order photon coherence $g^{(2)}(t)$ for arbitrary classical or quantum input states is the central theoretical result of this work. At coincidence $t=0$ [Eq. (\ref{eq:g2withpolarizationAppendix})], one recovers Eq. (\ref{eq:g2withpolarization}) presented in the main text.

\subsection{$g^{(2)}(t=0)$ for Quasi-1D Systems}

In quasi-one-dimensional quantum materials composed of atomic chains, only photons linearly polarized along the chain direction (``$\hat{x}$'') can efficiently couple to the material's elementary excitations. Suppose now that the input field is an arbitrary classical or quantum drive, but is linearly polarized along the material's active axis $\hat{x}$. In this case, polarization can be ignored, and the output can be obtained via a single trajectory $j = 0$. The one-photon ($\expect{ \BinXD{x} \BinX{x} }_{L,\wIn}$) and two-photon ($\expect{ \BinXD{x} \BinXD{x} \BinX{x} \BinX{x} }_{L,\wIn}$) input field correlation functions trivially determine the one- and two-photon input state decomposition, reading
\begin{align}
    \fVec{j=0} &= \sqrt{ \expect{ \BinXD{x} \BinX{x} }_{L,\wIn} } ~~ \polVec_{x} \\
    \FVecIJ{i=0}{j=0} &= \sqrt{ \expect{ \BinXD{x} \BinXD{x} \BinX{x} \BinX{x} }_{L,\wIn} } ~~ \polVec_{x}
\end{align}
Similarly, since only cavity photons with polarization $\polVec_x$ couple to the material, the sole non-vanishing component of the polarization-space Raman spin-photon operator is
\begin{align}
    \hat{D}_{xx} \equiv \hat{D} ~=~ \polVecC_x \cdot \Dop \cdot \polVec_x
\end{align}
Consequently, a simplified formula for $g^{(2)}(t)$ for one-dimensional systems and $x$-polarized photons can be derived by replacing the full spin-photon operator operator as $\Dop = \hat{D} \otimes \polVec_x \cdot \polVecC_x$, where the outer product projects photon polarizations onto $x$.

Substitution into Eq. (\ref{eq:G1complete}), the one-photon output correlation function  for one-dimensional systems becomes
\begin{align}
    G^{(1)}(t) = \gamma^2 \sum_{n} &\left| \bra{n} \frac{1}{\wIO + E_0 - \Ham - \hat{D} + i\gamma} \ket{0} \right|^2 \times e^{i(\wIn + E_0 - E_n) t} ~ \expect{ \BinXD{x} \BinX{x} }_{L,\wIn}  \label{eq:G1_1D_appendix}
\end{align}
Similarly, substitution into Eq. (\ref{eq:G2complete}) yields the two-photon output correlation function
\begin{align}
    &G^{(2)}(t) = \gamma^4 \sum_{n} \left| \bra{n} \left\{ \sum_m e^{i (E_n - E_m) t} \left[ \frac{1 - e^{i(\wIO+E_m-\Ham-\hat{D}+i\gamma) t}}{\wIO + E_m - \Ham - \hat{D} + i\gamma} \right] \ketbra{m}{m} \left[ \frac{1}{\wIO + E_0 - \Ham - \hat{D} + i\gamma} \right] \right.\right. \notag\\
    &+ \left.\left. \left[ e^{i(\wIO + E_n - \Ham - \hat{D} + i\gamma)t} \right]
    \frac{2}{2\wIO + E_0 - \Ham - 2 \hat{D} + 2i\gamma} ~\frac{1}{\wIO + E_0 - \Ham - \hat{D} + i\gamma}  \right\} \ket{0} \right|^2 ~~\expect{ \BinXD{x} \BinXD{x} \BinX{x} \BinX{x} }_{L,\wIn}  \label{eq:G2_2D_appendix}
\end{align}
The second-order photon coherence for one-dimensional systems can be obtained via
\begin{align}
    g^{(2)}(t) = \frac{G^{(2)}(t)}{\left[ G^{(1)}(0) \right]^2}
\end{align}
At coincidence $t=0$, process (I) again vanishes. The second order photon coincidence therefore recovers Eq. (\ref{eq:g2_1D}) provided in the main text. The inset depicting $g^{(2)}(t > 0)$ in Fig. 2 is computed numerically via Eqs. (\ref{eq:G1_1D_appendix}) and (\ref{eq:G2_2D_appendix}) at finite time delay $t$.

Notably, for numerical evaluation, the second-order photon coincidence can be straightforwardly rewritten as
\begin{align}
    g^{(2)}(t=0) = \frac{2 \Imag \left< \frac{1}{\wIO + E_0 - \Ham - \hat{D} - i\gamma} \frac{1}{2\wIO + E_0 - \Ham - 2\hat{D} + 2i\gamma} \frac{1}{\wIO + E_0 - \Ham - \hat{D} + i\gamma} \right>}{\left[ \Imag \left< \frac{1}{\wIO + E_0 - \Ham - \hat{D} + i\gamma} \right> \right]^2}
\end{align}
Here, the denominator can be computed using standard Lanczos or kernel polynomial methods. The numerator can be computed via applying the resolvent $\ket{\phi} = [ \wIO + E_0 - \Ham - \hat{D} + i\gamma]^{-1} \ket{0}$ to the material ground state using, e.g., conjugate gradient expansions, then again computing the expectation value $\bra{\phi} [ 2\wIO + E_0 - \Ham - 2\hat{D} + 2i\gamma ]^{-1} \ket{\phi}$ using, e.g., Lanczos iteration.

\end{widetext}

\section{Scattering Matrix Approach to the Second-Order Photon Coherence}
\label{app:scatteringMatrixApproach}

This section provides a complementary derivation of the second-order (pair) output photon correlation function $G^{(2)}(t)$ for a single photon polarization state [obtained above using input-output theory in Eq. (\ref{eq:G2_2D_appendix})] using a scattering matrix approach. This computation however needs to be phenomenologically modified with a non-Hermitian term for the cavity Hamiltonian, to account for the finite cavity photon lifetime. We treat the amplitude $\sqrt{\gamma}$ for photons tunneling through the cavity mirrors perturbatively. Importantly, we however again treat the scattering of photons off the material non-perturbatively. $G^{(2)}(t)$ thereby follows from an infinite-order sequence of Raman scattering processes inside the cavity before the photons tunnel out, encoding the non-perturbative dressing of matter with light. We start by separating the full Hamiltonian into a part $\Ham$ describing the material ($\Ham_0$) and cavity photon modes as well as light-matter coupling within the cavity, and a perturbation $\hat{V}$ that describes photon tunneling through the mirrors. Ignoring photon polarization (i.e., assuming a quasi-one-dimensional material), these contributions read:
\begin{align}
	\Ham &= \Ham_0 + \left(\omega_0 + \hat{D}  \right) \AD{} \A{} + \sum_{\alpha=L,R} \int d\omega~ \omega~ \BD{\alpha}(\omega) \B{\alpha}(\omega) \\
	\hat{V} &= \sqrt{\frac{\gamma}{2\pi}} \sum_{\alpha=L,R} \int d\omega~ \left[ \AD{} \B{\alpha}(\omega) + \hc \right]
\end{align}
Using the interaction-picture time evolution operator $\hat{U}(t,t') = \mathcal{T} \exp\{ \int_{t'}^{t} d\tau V(\tau) \}$, the pair-photon output correlation function can be computed as a transition rate
\begin{align}
	G^{(2)}(t) = \sum_f \left| \bra{f(t)} e^{-i \Ham_0 t_f} \hat{U}(t_f,t_i) e^{i \Ham_0 t_i} \ket{i} \right|^2
\end{align}
from an initial state
\begin{align}
    \ket{i} &= \ket{G}_{\rm cav} \otimes \ket{\rm in}_{L,\wIn} \otimes \ket{0}_R
\end{align}
at time $t_i \to -\infty$ (where $\ket{G}_{\rm cav}$ denotes the cavity and material in its ground state, $\ket{0}_R$ denotes the right-side free-space modes in their vacuum state, and $\ket{\rm in}_{L,\wIn}$ is a left-side monochromatic input field with frequency $\wIn$) to a final state
\begin{align}
    \ket{f(t)} &= \ket{f}_{\rm cav} \otimes \ket{f}_{L} \otimes \left[ \BoutXD{R}(t_m) \BoutXD{R}(t_m+t) \ket{0}_R \right]
\end{align}
at time $t_f \to +\infty$ which comprises final states for the material and cavity as well as left-side photon fields, and wave packets of two output photons detected on the right, with a time delay $t$. For the stationary input fields considered here, the result will be independent of the time $t_m$ of observing the first photon. Following Ref. \cite{gardiner85}, the output fields are defined as Fourier transforms of the free-space mode continuum $\BoutX{R}(t) = \frac{1}{\sqrt{2\pi}} \int d\omega~ \B{R}(\omega) e^{-i\omega(t-t_f)}$. The final state $\ket{f}$ is therefore \textit{not} an eigenstate of the right-side photon bath, but involves two frequency integrals over the right-side photon modes. 

To lowest (fourth) order in $\sqrt{\gamma}$, two processes processes contribute to the pair photon correlation function: (I) the sequential tunneling of two photons through the cavity, and (II) the tunneling of two photons into the cavity (yielding a two-photon intermediate state), followed by their sequential tunneling out. The final state has two fewer photons (with energy $\wIn$) on the left and two photons (with net energy $\Omega$) on the right. The energy difference between final and initial states therefore reads $\Omega - 2\wIn + E_f - E_0$, where $E_f$ and $E_0$ are the final-state and ground state energies for the material, respectively. The pair photon correlation function therefore reads
\begin{align}
    &G^{(2)}(t) = \sum_f \left| \int \frac{d\omega d\Omega}{2\pi} e^{-i[\Omega(t_m-t_f) + \omega t]}  \right. \notag\\
    &~~\times \left. \bra{1_{\omega},1_{\Omega-\omega}}_R \bra{f}_{\textrm{cav},L} \frac{1}{2\wIn + E_0 - E_f - \Omega + i\eta} \hat{T} \ket{i} \right|^2  \label{eq:G2FGR}
\end{align}
where $\hat{T}$ is defined to fourth order in photon tunneling through the mirrors as
\begin{align}
    \hat{T} = \hat{V} \frac{1}{E_i - \Ham + i\eta} \hat{V} \frac{1}{E_i - \Ham + i\eta} \hat{V} \frac{1}{E_i - \Ham + i\eta} \hat{V}
\end{align}
with $\eta$ is a positive convergence factor. To find an explicit expression for $\hat{T}$, one can simply count photon numbers and photon energies. For instance, the right-most tunneling operator $\hat{V}$ must remove a photon from the monochromatic input field (yielding an energy $\wIn$) and add it to the cavity (costing an energy $\omega_0$). The right-most resolvent must therefore read $[\wIn - \omega_0 + E_0 - (\Ham_0 - \hat{D}) + i\eta]^{-1}$, where $E_0$ is the material ground state energy. Furthermore, $[E_i - E_f + i\eta]^{-1} = [2\wIn - \Omega - \Delta + i\eta]^{-1}$ where $\Delta$ is the energy deposited into the material.
\begin{widetext}
Using $\wIO \equiv \wIn - \omega_0$ the detuning of the input photon from the bare cavity resonance, the sequential tunneling process (I) can now be expanded to
\begin{align}
    \hat{T}^{(I)} &= \frac{\gamma^2}{(2\pi)^2} \BD{R}(\Omega-\omega) \frac{1}{\wIO + \wIn - \omega + E_0 - \Ham_0 - \hat{D} + i\eta} \B{L} \frac{1}{\wIn - \omega + E_0 - \Ham_0 + i\eta} \BD{R}(\omega) \frac{1}{\wIO + E_0 - \Ham_0 - \hat{D} + i\eta} \B{L} \notag\\
        &+ [\omega \to (\Omega-\omega)] \\
    \hat{T}^{(II)} &= \frac{\gamma^2}{(2\pi)^2} \BD{R}(\Omega-\omega) \frac{1}{\wIO + \wIn - \omega + E_0 - \Ham_0 - \hat{D} + i\eta} \B{R}(\omega) \frac{1}{2\wIO + E_0 - \Ham_0 - 2\hat{D} + i\eta}  \B{L} \frac{1}{\wIO + E_0 - \Ham_0 - \hat{D} + i\eta} \B{L} \notag\\
        &+ [\omega \to (\Omega-\omega)]
\end{align}
These processes jointly compose a perturbative expression for the observed pair photon correlation function:
\begin{align}
    G^{(2)}(t) = \gamma^4 \sum_f \left| \frac{1}{(2\pi)} \bra{f}_{\rm cav} \left[ \hat{M}^{(I)}(t) + \hat{M}^{(II)}(t) \right] \ket{G}_{\rm cav} \right|^2 \expect{ \BD{L} \BD{L} \B{L} \B{L} }_{L,\wIn}   \label{eq:G2FGRtwoparts}
\end{align}
where $\B{L}$ is short-hand for removing an input photon with frequency $\wIn$ on the left side. First, consider process (II):
\begin{align}
    \hat{M}^{(II)}(t) &= 2 \int \frac{d\omega d\Omega}{2\pi} e^{i\Omega (t_f - t_m)} \left[ e^{-i\omega t} + e^{i(\omega-\Omega)t} \right] \frac{1}{2\wIn + E_0 - \Omega - E_f + i \eta} \frac{1}{\wIO + \wIn - \omega + E_0 - \Ham_0 - \hat{D} + i\eta} \notag\\
        &\cdot \frac{1}{2\wIO + E_0 - \Ham_0 - 2\hat{D} + i\eta} \frac{1}{\wIO + E_0 - \Ham_0 - \hat{D} + i\eta}
\end{align}
The prefactor $2$ originates from evaluating the expectation value for the integrated-out cavity photons $\expect{ \A{} \A{} \AD{} \AD{} } = 2$.
The final-state frequency integrals can be evaluated using $(1/2\pi)\int d\omega~ e^{i \omega t}/(A - \omega + i\eta) = -i e^{i A t} \theta(t)$ with the step function $\theta$. Assuming a positive time delay $t>0$ and $t_f \to +\infty$, one finds
\begin{align}
    \hat{M}^{(II)}(t) &= -4\pi e^{i(2\wIn + E_0 - E_f)(t_f-t_m-t)} e^{i(\wIO + \wIn + E_0 - \Ham_0 - \hat{D})t}  \frac{1}{2\wIO + E_0 - \Ham_0 - 2\hat{D} + i\eta} \frac{1}{\wIO + E_0 - \Ham_0 - \hat{D} + i\eta}
\end{align} 
For process (I), a complete set of intermediate states $\Ham_0 \ket{m} = E_m \ket{m}$ can be inserted after the first photon tunnels through the cavity, to perform the frequency integrals:
\begin{align}
    \hat{M}^{(I)}(t) &= \int \frac{d\omega d\Omega}{2\pi} e^{i\Omega (t_f - t_m)} \left[ e^{-i\omega t} + e^{i(\omega-\Omega)t} \right] \frac{1}{2\wIn + E_0 - \Omega - E_f + i \eta} \frac{1}{\wIO + \wIn - \omega + E_0 - \Ham_0 - \hat{D} + i\eta} \notag\\
        &\cdot \sum_m \ketbra{m}{m} \frac{1}{\wIn - \omega + E_0 - E_m + i\eta} \frac{1}{\wIO + E_0 - \Ham_0 - \hat{D} + i\eta}
\end{align}
Using $(1/2\pi) \int d\omega~ e^{i\omega t} / [(\hat{A} - \omega + i\eta)(E - \omega + i\eta)] = i \theta(t) ( e^{i\hat{A}t} - e^{iEt} )/(\hat{A}-E)$, one finds (assuming $t>0$):
\begin{align}
    \hat{M}^{(I)}(t) &= 2\pi \sum_m e^{i(2\wIn + E_0 - E_f)(t_f-t_m-t)} e^{i(\wIn+E_0-E_m) t} \frac{e^{i(\wIO + E_m - \Ham_0 - \hat{D})t} - 1}{\wIO + E_m - \Ham_0 - \hat{D}} \ketbra{m}{m} \frac{1}{\wIO + E_0 - \Ham_0 - \hat{D} + i\eta}
\end{align}
Combining both expressions with Eq. (\ref{eq:G2FGRtwoparts}), and factoring out an overall phase factor $e^{i(2\wIn + E_0 - E_f)(t_f-t_m)} e^{-i\wIn t}$, we arrive at
\begin{align}
    G^{(2)}(t) &= \sum_f \left| \gamma^2 \bra{f}_{\rm cav} \left[ 2 e^{i(\wIO + E_f - \Ham_0 - \hat{D})t}  \frac{1}{2\wIO + E_0 - \Ham_0 - 2\hat{D} + i\eta} \frac{1}{\wIO + E_0 - \Ham_0 - \hat{D} + i\eta} \right.\right. \notag\\
    & \left.\left. - \sum_m e^{i(E_f-E_m) t} \frac{e^{i(\wIO + E_m - \Ham_0 - \hat{D})t} - 1}{\wIO + E_m - \Ham_0 - \hat{D}} \ketbra{m}{m} \frac{1}{\wIO + E_0 - \Ham_0 - \hat{D} + i\eta}  \right] \ket{G}_{\rm cav} \right|^2 \expect{ \BD{L} \BD{L} \B{L} \B{L} }_{L,\wIn}
\end{align}
This \textit{almost} recovers the pair-photon output correlation function obtained using input-output theory above [Eq. (\ref{eq:G2_2D_appendix}]. If photon tunneling into/out of the cavity is treated perturbative, all intermediate intra-cavity one- and two-photon states have infinite lifetime (as $\eta$ is merely an infinitesimal convergence factor). To recover the proper dependence on the finite cavity photon decay rate $\gamma$, one must now \textit{a posteriori} replace $\eta \to \gamma$ in the equations above. This is equivalent to introducing a non-Hermitian contribution to the intra-cavity Hamiltonian, via replacing $\omega_0 \AD{} \A{} \to (\omega_0 + i\gamma) \AD{} \A{}$. We note that the dependence on the broadening $\gamma$ could be rigorously recovered by resumming an appropriate class of infinite-order (in $\sqrt{\gamma}$) photon tunneling processes; these are automatically accounted for when combining input-output relations with a non-perturative solution of the cavity + material Langevin equation described in the previous section.

\end{widetext}

\section{TFIM Simulations}
\label{app:TFIMsims}

Calculations for the TFIM were performed for a 256 site chain with periodic boundary conditions (PBC). Although the eigenstates are exactly solvable through a Jordan-Wigner transformation, the non-perturbative $g^{(2)}(0)$ formula requires all excitations and therefore some approximations as detailed below.

For an $N$-site chain (with $N$ even), the Hamiltonian separates into $N/2$ momentum sectors. For PBC, the momenta take the values $k=2\pi(l-1/2)/N$ for $l=\left[1,2,...,N/2\right]$, chosen so that $0<k<\pi$. Within each momentum sector, we have

\begin{align}
    \Ham_l &= -E_l \sigma_z \\
    \hat{D}_l &= D\cos{(k_l-\theta_l)}\sigma_z + D\sin{(k_l-\theta_l)}\sigma_x,
\end{align}

where $\sigma_{x,z}$ are Pauli matrices and

\begin{align}
    E_l &= \sqrt{J^2+h^2+2Jh\cos{k_l}}  \\
    \theta_l &= \cos^{-1}\left(\frac{h+J\cos{k_l}}{E_l}\right).
\end{align}

Using these formulas, we calculate the eigenstates of $\Ham_l+\hat{D}_l$, which we write as $\ket{n_l}_{l}^{(1)}$ for $n_l=0,1$, with energy $E^{(1)}_{n_l,l}$. From these, we can write eigenstates of $H+D$ as $\ket{n}^{(1)}=\bigotimes_l\ket{n_l}_{l}^{(1)}$ with energy $E^{(1)}_{n}=\sum_l E^{(1)}_{n_l,l}$. We express the energies in terms of excitations $E^{(1)}_{n}=E^{(1)}_{0}+\sum_l \Delta E^{(1)}_{n_l,l}$. We repeat this process for $\Ham_l+2\hat{D}_l$ written with $(1)\to(2)$.

We also compute the overlaps in eigenstates for each momentum sector. We write the transition from the zero photon ground state to an arbitrary one photon state as $T^{1,0}_{n_l,0,l}\equiv \leftindex^{(1)}_l{} \braket{n_l}{0}_l$. The transition from one photon to two photon states is $T_{m_l,n_l,l}^{2,1}\equiv\leftindex^{(2)}_l{}\braket{m_l}{n_l}_{l}^{(1)}$. 

The photon coincidence spectrum is then computed from Eq. (\ref{eq:g2_1D}) by inserting complete bases of eigenstates of $\Ham+\hat{D}$ and $\Ham+2\hat{D}$. For example, the $G^{1}$ term in the denominator becomes

\begin{align}
    &\sum\limits_n \left| \leftindex^{(1)}{}{\bra{n}} \frac{1}{\wIO + E_0 - \Ham - \hat{D} + i\gamma} \ket{0} \right|^2\notag\\
    &=\sum\limits_n \left| \frac{\leftindex^{(1)}{}{\braket{n}{0}}}{\wIO + E_0 - E^{(1)}_{n} + i\gamma}\right|^2\notag\\
    &=\sum\limits_n \left| \frac{\prod_l T_{n_l,0,l}^{1,0}}{\wIO + E_0 - E^{(1)}_{0} - \sum_l \Delta E^{(1)}_{n_l,l} + i\gamma}\right|^2
\end{align}

At this point, however, we need to make an approximation, since there are $2^{N/2}$ states $\ket{n}$ in the exact calculation. We construct an algorithm for excluding any terms found to contribute less than a tolerance $\delta$, which gets set by checking for convergence as the tolerance is decreased.

The goal of the algorithm is to identify states where small transition amplitudes in the numerator or large energies in the denominator lead to small contributions. For the numerator, we note that all transitions have $\left|T_l\right|\le1$. Therefore, if the contribution from the first few values of l is small, we know that the total product in the numerator can only get smaller. Similarly, the excitation energies in the denominator are all positive. Therefore, if the contribution from the first few values of l is larger than $\wIO + E_0 - E^{(1)}_{0}$, then the magnitude of the denominator can only get larger. Using these observations, we sweep through $l$, and on the $L^{th}$ step branch into the two states defined by $n_L=0,1$. For each, we compute an estimate for the best possible weight from the state so far

\begin{align}
    \frac{\prod_{l\le L}T_{n_l,0,l}^{1,0}}{\max\left(\sum_{l\le L}\Delta E^{(1)}_{n_l,l}+E^{(1)}_{0}-E_0-\wIO,\gamma\right)}. \label{eq:pruningtolerance}
\end{align}

Whenever this value drops below $\delta$, we immediately prune that branch. Note that for small system sizes, the procedure matches exact diagonalization when the tolerance $\delta\to0$.

The calculation for $G^2$ is complicated by needing to include transitions from one to two photons. One needs to take care to sum over the intermediate one-photon $\ket{n}$ states before squaring the total value for each two-photon $\ket{m}$ state, but otherwise the pruning procedure is the same.

\begin{figure}[h!]
    \centering
    \includegraphics[width=0.9\columnwidth]{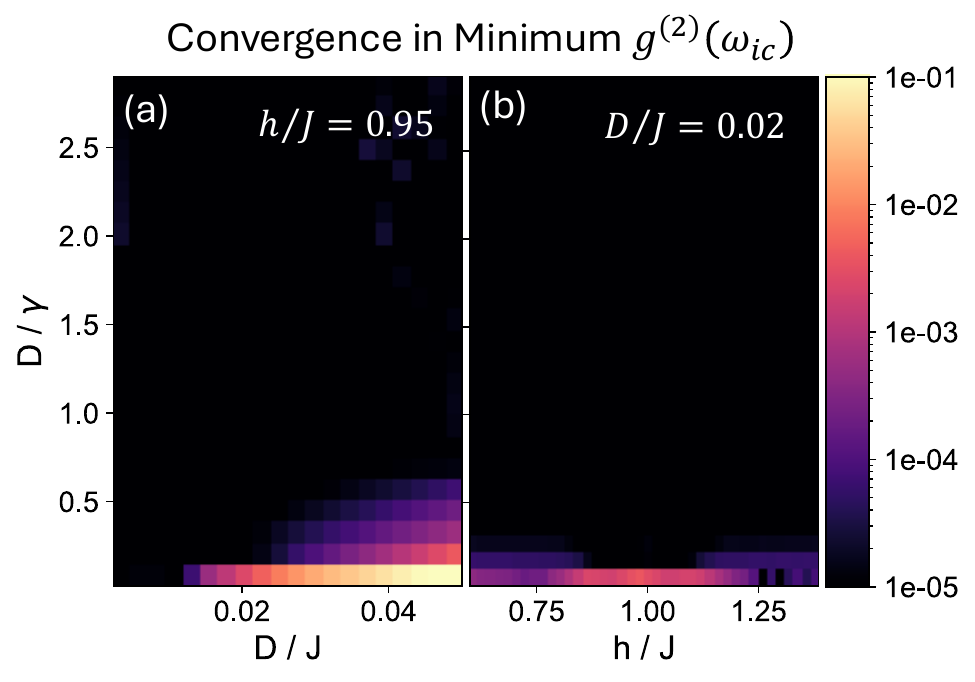}
    \caption{\textbf{TFIM Simulation Convergence.} The difference in the minimum value of $g^{(2)}$ computed with $\delta=10^{-4}$ and $\delta=10^{-5}$ for $h/J=0.95$ (a) and $D/J=0.02$ (b).}
    \label{fig:convergence}
\end{figure}

For our 256 site chain, tolerances of $\delta=10^{-5}$ (on the value in Eq. (\ref{eq:pruningtolerance}) before squaring) are easily achievable. Tolerances used in this paper range from $\delta=10^{-4}$ to $\delta=10^{-5}$, which can result in $\sim 10^7$ states being included. Figure \ref{fig:convergence} shows the change in the minimum value of $g^{(2)}$ calculated when using $\delta=10^{-4}$ and $\delta=10^{-5}$.

\section{EPR Pair Generation in a Minimal Model of a Chiral Material}
\label{app:EPRsims}

To illustrate the polarization-sensitive many-body photon blockade, we study a minimal toy model of a chiral magnetic system using exact diagonalization. We choose a single four-site plaquette, coupled via nearest-neighbor Ising exchange interactions $J$ and next-nearest-neighbor off-diagonal $\Gamma$ exchange terms. The Hamiltonian reads
\begin{align}
    \Ham = -J \sum_{\left<ij\right>} \hat{S}_i^x \hat{S}_j^x - \Gamma \sum_{\left<\left<ij\right>\right>} \left( \hat{S}_i^x \hat{S}_j^y + \hat{S}_i^y \hat{S}_j^x \right)
\end{align}
In principle, Raman spin-photon processes $\sim \hat{D}_{\mu\nu} \AD{\mu} \A{\nu}$ can tune all symmetry-allowed exchange interactions as a function of photon polarization. We therefore choose a minimal model that retains the salient symmetry-allowed features, reading:
\begin{align}
    \hat{D}_{xx} &= D \sum_i S_i^x S_{i + \hat{x}}^x \\
    \hat{D}_{yy} &= D \sum_i S_i^x S_{i + \hat{y}}^x \\
    \hat{D}_{xy} &= D \sum_i \left( S_i^+ S_{i + \hat{x} + \hat{y}}^+ - S_i^- S_{i - \hat{x} + \hat{y}}^- \right) \\
    \hat{D}_{yx} &= \hat{D}_{xy}^\dag = D \sum_i \left( S_i^- S_{i + \hat{x} + \hat{y}}^- - S_i^+ S_{i - \hat{x} + \hat{y}}^+ \right)
\end{align}
Under four-fold rotations, $C_4 \hat{D}_{xx} C_4^\dag = \hat{D}_{yy}$ and 
$C_4 \hat{D}_{xy} C_4^\dag = -\hat{D}_{yx}$. The joint light-matter-coupled Hamiltonian $\Ham_{\rm cav} = \Ham + \sum_{\mu\nu} \hat{D}_{\mu\nu} \AD{\mu} \A{\nu}$ therefore remains invariant under rotations, however mirror symmetry (requiring $\sigma_{\rm h} \hat{D}_{xy} \sigma_{\rm h}^\dag = -\hat{D}_{xy}$) is explicitly broken.

The polarization-selective second-order photon coherence $g_{\mu\mu'}^{(2)}(t)$ presented in the main text was computed using Eq. (\ref{eq:g2withpolarizationAppendix}), using $\Gamma/J = 0.6$ and $D/J = 0.1$ for a four-site plaquette.

\clearpage

\end{document}